# A secondary atmosphere on the rocky exoplanet 55 Cancri e

Renyu Hu[1,2], Aaron Bello-Arufe[1], Michael Zhang[3], Kimberly Paragas[2], Mantas Zilinskas[4], Christiaan van Buchem[5], Michael Bess[6], Jayshil Patel[7], Yuichi Ito[8,9], Mario Damiano[1], Markus Scheucher[1], Apurva V. Oza[1], Heather A. Knutson[2], Yamila Miguel[4,5], Diana Dragomir[6], Alexis Brandeker[7], Brice-Olivier Demory[10,11]

1. Jet Propulsion Laboratory, California Institute of Technology, Pasadena, CA 91011, USA
2. Division of Geological and Planetary Sciences, California Institute of Technology, Pasadena, CA 91125, USA
3. Department of Astronomy and Astrophysics, The University of Chicago, Chicago, IL 60637, USA
4. SRON Netherlands Institute for Space Research, Leiden, The Netherlands
5. Leiden Observatory, Leiden University, Leiden, The Netherlands
6. Department of Physics and Astronomy, University of New Mexico, Albuquerque, NM 87131, USA
7. Department of Astronomy, Stockholm University, Stockholm 10691, Sweden
8. Division of Science, National Astronomical Observatory of Japan, 2-21-1 Osawa, Mitaka, Tokyo, 181-8588, Japan
9. Department of Physics and Astronomy, University College London, Gower Street, WC1E 6BT London, United Kingdom
10. Center for Space and Habitability, University of Bern, Bern 3012, Switzerland
11. Space Research and Planetary Sciences, Physics Institute, University of Bern, Bern 3012, Switzerland

Contact: renyu.hu@jpl.nasa.gov

**Summary Paragraph**

Characterizing rocky exoplanets is a central endeavor of astronomy, and yet the search for atmospheres on rocky exoplanets has hitherto resulted in either tight upper limits on the atmospheric mass[1–3] or inconclusive results[4–6]. The 1.95-$R_{Earth}$ and 8.8-$M_{Earth}$ planet 55 Cnc e, with a predominantly rocky composition and an equilibrium temperature of ~2000 K, may have a volatile envelope (containing molecules made from a combination of C, H, O, N, S, and P elements) that accounts for up to a few percent of its radius[7–13]. The planet has been observed extensively with transmission spectroscopy[14–22], and its thermal emission has been measured in broad photometric bands[23–26]. These observations disfavor a primordial $H_2$/He-dominated atmosphere, but cannot conclusively determine whether the planet has a secondary atmosphere[27,28]. Here we report a thermal emission spectrum of the planet obtained by JWST's NIRCam and MIRI instruments from 4 to 12 μm. The measurements rule out the scenario where the planet is a lava world shrouded by a tenuous atmosphere made of vaporized rock[29–32], and indicate a bona fide volatile atmosphere likely rich in $CO_2$ or CO. This atmosphere can be outgassed from and sustained by a magma ocean.

**Main text**

Using JWST, we observed the secondary eclipse of 55 Cnc e (i.e., when the planet passes behind the star) using NIRCam grism time series with the F444W filter on November 24, 2022, and another secondary eclipse with MIRI low-resolution spectroscopy on March 24, 2023. Due to the brightness of the star ($m_K$=4.0), we used 2 groups per integration for the NIRCam observation and 5 groups per integration for the MIRI observation. None of the integrations of the NIRCam observation are saturated, and the pixels on the MIRI detector for wavelengths between 5.0-6.0 um are saturated at the 2nd group (i.e., fully saturated) and those for wavelength between 6.0-7.7 um are saturated between the 3rd and the 5th group (i.e., partially saturated). We ignored the fully saturated pixels but made efforts to recover the partially saturated pixels.

The NIRCam data were analyzed by 4 independent teams with different pipelines (Methods). We removed 1/f noise (the correlated read noise that impacts data across a wide range of timescales with a 1/f power spectrum[33]) with a row-by-row subtraction based on the unilluminated part of the detector and found the procedure greatly reduced the scatter of the light curve. The extracted white light curves of the NIRCam observation have a substantial decreasing trend, and we applied various ways to correct for this trend in the independent analyses, including the typically used exponential and linear ramps, detrending against the focal plane array housing temperature, and with or without detrending against the spectral trace positions on the detector. All methods were able to effectively remove the decreasing trend, but substantial correlated noise (on the order of 100 ppm and with a semi-periodic time scale of 1 hour) remained in the light curve (Fig. 1). We could not determine the origin of this correlated noise, which could be unknown instrument systematics or astrophysical noise (e.g., stellar supergranulation[34]). We divided the spectroscopic light curves by the white light curve and found that the correlated noise was significantly reduced, implying its achromatic nature (Extended Data Fig. 1). Based on the relative spectroscopic light curves, we derived a relative eclipse depth spectrum and then allowed an arbitrary absolute eclipse depth in the interpretation (Fig. 2). The shape of the relative eclipse depth spectrum is consistent with the spectral shape from direct-fitting to the spectroscopic light curves, minus an offset, indicating the validity of our approach (Extended Data Fig. 1).

The MIRI data were analyzed by 2 independent teams, notably with different ways to recover the partially saturated pixels and correct for detector nonlinearity (Methods). We found that the spectral trace on the detector shifted substantially in the spatial and spectral directions nearly coincidentally with the eclipse, and without correcting for the drift, the derived eclipse depth would be nearly zero. With a linear decorrelation against this drift, the eclipse is recovered by both analyses. To ensure that the decorrelation was not biased by any correlation between the eclipse and the drift, we also estimated the decorrelation coefficients after masking the data taken during the eclipse and found that they were consistent with the values derived from the full dataset. The two analyses produced consistent results (Extended Data Fig. 2), and the level of correlated noise in the MIRI white light curves is significantly smaller than that of NIRCam, allowing for precise measurement of the eclipse depth (Fig. 1) and spectrum (Fig. 2).

Based on the white light curve at 6.3-11.8 um (integrating the spectral channels that do not contain fully saturated pixels), we measured an eclipse depth of $110^{+8}_{-9}$ ppm. The eclipse depth excluding the shadow region from 10.5 to 11.8 µm[35] is $112^{+8}_{-9}$ ppm. Using the stellar spectrum also measured in the same observation, we obtained a brightness temperature of 1796±88 K. This brightness temperature is by 8σ lower than the temperature for zero albedo and zero heat redistribution (2511±26 K), which is expected and typically measured for airless rocky exoplanets in the thermal infrared[1–3,6]. Factoring in a 10% stellar flux measurement bias could reduce the discrepancy to 6.6σ. A high-albedo surface (Bond albedo>0.5) could make the brightness temperature consistent with the measured value within 3σ, however, such a high-albedo surface is inconsistent with the albedo of molten silicates[36] or TESS and CHEOPS observations of the planet in visible light[37–40]. Realistic scenarios for a molten lava surface, which would include a tenuous rock-vapor atmosphere in equilibrium[31,32], have eclipse depths in MIRI wavelengths that far exceed the measurements (Fig. 3; Extended Data Fig. 7). The brightness temperature measured thus indicates that there must be heat redistribution occurring on the planet, consistent with the indication of prior Spitzer observations[23,27]. The molten lava cannot provide effective heat redistribution on a planet as hot as 55 Cnc e[41], and the heat redistribution is most readily provided by a substantial atmosphere[42].

We fit the thermal emission spectrum of 55 Cnc e in 4-12 µm with a Bayesian retrieval framework that statistically samples a wide range of pressure-temperature profiles and atmospheric molecular composition (Methods). We allowed the absolute eclipse depth in the NIRCam bandpass to vary freely, accounting for not only the correlated noise in the NIRCam observation but also possible variations in the overall planet brightness over the 4 months between the NIRCam and MIRI observations. We first fit the data with a blackbody to establish the null-hypothesis baseline, and then carried out model fits that include $H_2O$, $CO$, $CO_2$, $N_2$, $O_2$, $SO_2$, or $SiO$ as a single-gas or binary-mixture composition to determine the gain in Bayes evidence (Methods). Among these, using $CO_2$ as a single-gas composition (for varied pressure-temperature profiles and surface pressures) and binary mixtures between $CO_2$, $CO$, and $N_2$ result in models that are preferred over the null hypothesis by >3σ when using the NIRCam dataset only, and >2σ when using the NIRCam and MIRI datasets together (Methods). Excluding the MIRI data points from the shadow region does not substantially affect this preference. The best-fit models with a CO or $N_2$ background atmosphere indicate an atmosphere of $10^3$-$10^7$ Pa (or 0.01-100 bar) that contains >$10^{-5}$ $CO_2$ by volume mixing ratio (Fig. 2; Extended Data Fig. 5). The correlation between the volume mixing ratio of $CO_2$ and the surface pressure indicates that the spectral modulation observed in 4-5 µm constrains the column density of $CO_2$ in these models. The posterior distribution of $CO_2$ has a tail towards small mixing ratios if CO is assumed as the background gas, indicating that a CO-only scenario is also allowed. Taken together, the spectral feature observed in 4-5 µm is best described by the absorption of $CO_2$ or CO in the planet's atmosphere.

To further study the range of physically plausible atmospheres on the planet, we calculated two sets of self-consistent atmospheric models and compared them with the measured spectrum. In one set of models, we simulated ~3500 self-consistent atmospheric chemistry and radiative transfer models[43] to explore varied CHONSP abundances in an assumed 200-bar atmosphere. The best-fit models center around $CO_2$-rich atmospheres (regardless of the background component) and CO-dominated atmospheres, consistent with the retrieval results (Fig. 3; Extended Data Fig. 8). The presence of $H_2O$, $SO_2$, or $PH_3$ could improve the fit to the spectral modulation in 4-5 μm in some cases. In the other set of models, we assumed an atmosphere in volatile equilibrium with the underlying magma ocean[44,45]. The overall C-H-N-S content of the atmosphere-magma system and the redox condition of the magma then control the atmospheric size and composition[44]. Despite the intense stellar irradiation and fast hydrodynamic escape, the atmosphere buffered by the magma ocean is sustained by the overall volatile inventory of the planet, which is hard to be completely removed by atmospheric escape (see Methods). With an Earth-like C-H-N-S abundance, we found that the resulting emission spectra can readily fit the data (Fig. 3). The magma ocean of a wide range of redox conditions maintains a CO-rich atmosphere with a $CO_2$ mixing ratio ranging from $10^{-1}$ to $10^{-4}$ (Extended Data Fig. 9), remarkably consistent with the preferred scenarios by spectral retrievals.

A secondary atmosphere can also explain the variability in the planet's thermal emission suggested by the earlier photometric measurements by Spitzer in the 4.5 μm band[24,25]. While this variability can be attributed to the formation and dissipation of a transient atmosphere[46], a change in the composition of the atmosphere (e.g., heating of the upper atmosphere due to injection of short-wave absorbers such as Na, K, Mg, MgO, SiO) can also cause the pressure-temperature profile to become more isothermal or even inverted[47] (Fig. 3; Extended Data Fig. 8). As a consequence, the absolute eclipse depth could change, and the absorption features in 4-5 μm can vanish or even become an emission feature over short timescales. Also, short-lived clouds fueled by condensable materials released from the magma ocean could temporarily raise the infrared photosphere and decrease the planet's thermal emission. Future observations with JWST and other observatories will help further understand the atmosphere and its interaction with the surface and interior of this intriguing rocky planet.

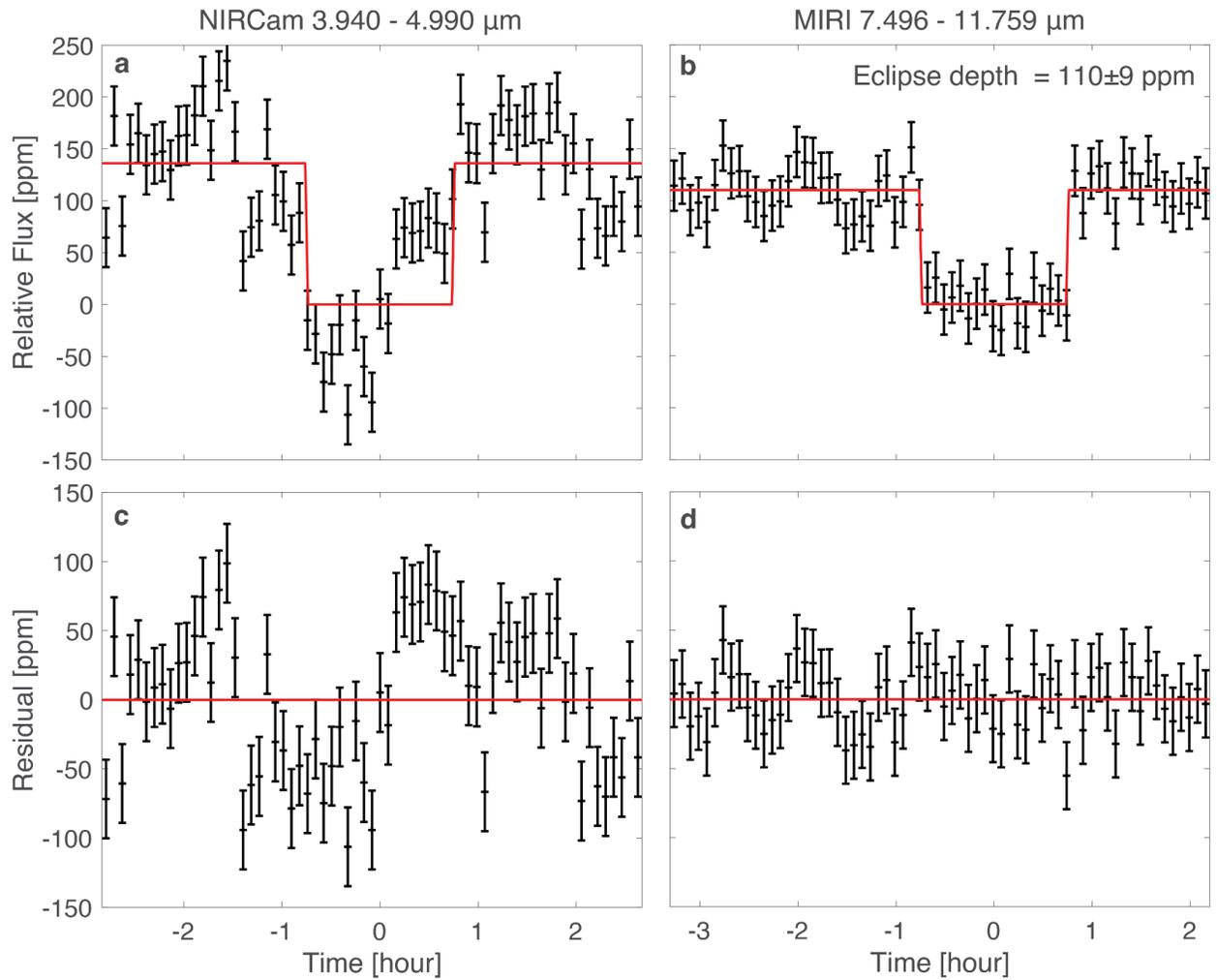

**Fig. 1: White light curves of the secondary eclipses of 55 Cnc e. a,** Detrended light curves of NIRCam observations. **b,** Detrended light curves of MIRI observations. **c, d,** Residuals of the light curves fitting to an eclipse model. The error bars correspond to 1σ.

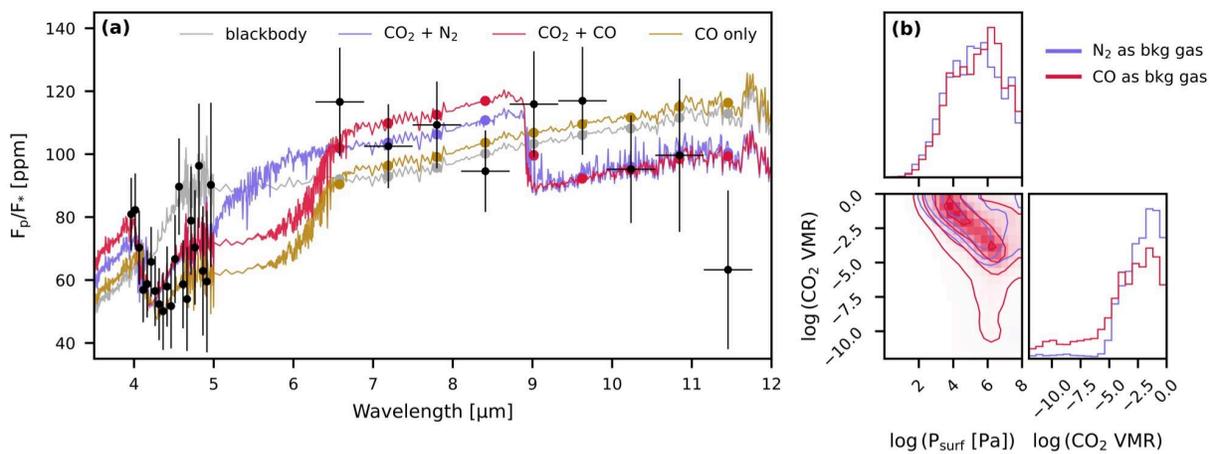

**Fig. 2: Thermal emission spectrum of 55 Cnc e. a,** Thermal emission spectrum of 55 Cnc e, overlying the best-fit models if the planet's spectrum is a blackbody, or if the planet has a $CO_2$-$N_2$, $CO_2$-CO, or CO-only atmosphere with varied composition and pressure-temperature profiles. Model results binned to the same wavelength channels as the MIRI data are shown as colored points. The error bars correspond to 1σ. **b,** Posterior distributions of spectral retrievals assuming $N_2$ (blue) or CO (red) as the background gas and the volume mixing ratio of $CO_2$ as a free parameter. The full posterior distributions and sample pressure-temperature profiles are shown in Extended Data Fig. 5.

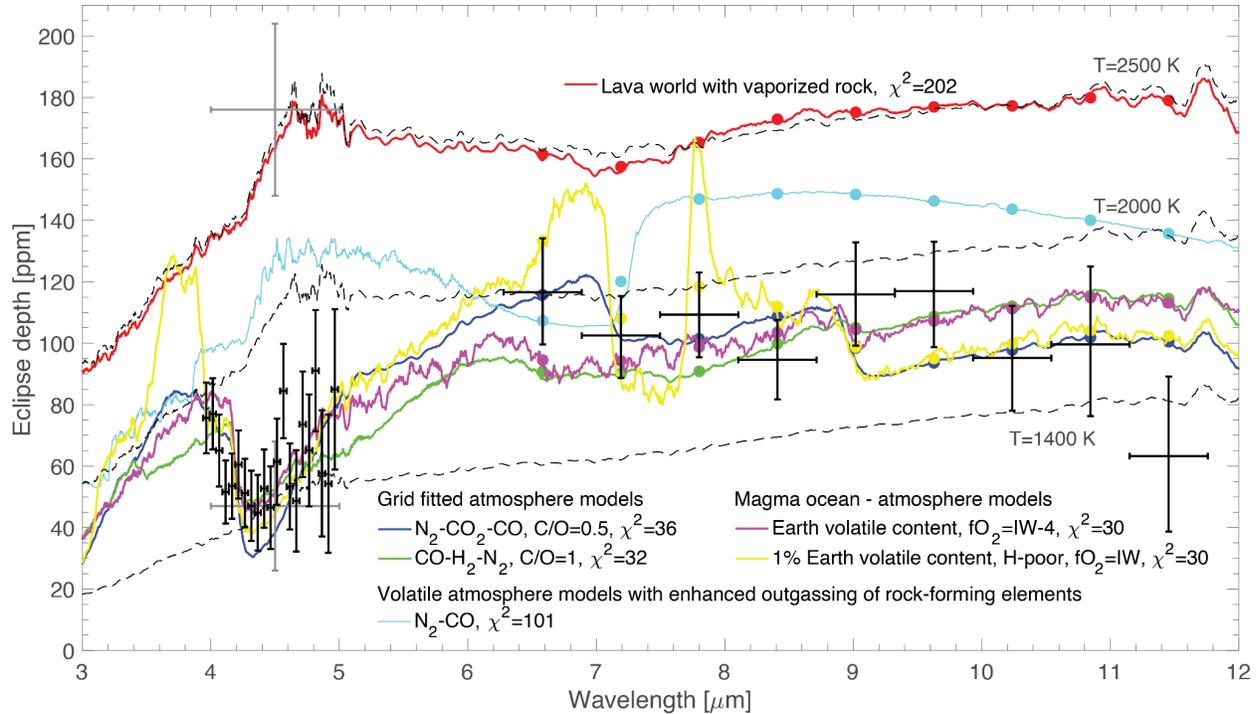

**Fig. 3: Self-consistent atmosphere models of 55 Cnc e.** The NIRCam relative eclipse depths are offset to a mean depth of 60 ppm. Previous Spitzer measurements[24] are shown in gray. The black dashed lines are blackbody models at the temperatures indicated in the plot. The vaporized-rock model was simulated using the rocky composition that corresponds to the bulk silicate Earth; using other plausible rock compositions results in qualitatively similar results (Extended Data Fig. 7). The atmosphere models are simulated either in a parameter exploration of CHONSP elemental abundances or with the assumption of volatile equilibrium with an underlying magma ocean. The models shown are selected from those that provide adequate fits to the data. The volatile atmosphere models with enhanced outgassing of rock-forming elements can cause substantial changes in the eclipse depth and the emission spectra. The models are smoothed to R=100 for the clarity of illustration. Model results binned to the same wavelength channels as the MIRI data are shown as colored points. The oxygen fugacity ($fO_2$) describes the redox condition of the magma and $fO_2$=IW-4 means a condition more reduced than the iron-wustite (IW) buffer by 4 orders of magnitude. The full sets of atmospheric models are described in Methods and Extended Data Figs. 8 and 9.

**Reference**

1. Kreidberg, L. *et al.* Absence of a thick atmosphere on the terrestrial exoplanet LHS 3844b. *Nature* **573**, 87–90 (2019).

2. Crossfield, I. J. *et al.* GJ 1252b: a hot terrestrial super-Earth with no atmosphere. *Astrophys. J. Lett.* **937**, 17 (2022).

3. Greene, T. P. *et al.* Thermal emission from the Earth-sized exoplanet TRAPPIST-1 b using JWST. *Nature* **618**, 39–42 (2023).

4. Tsiaras, A. *et al.* Detection of an atmosphere around the super-Earth 55 Cancri e. *Astrophys. J.* **820**, 99 (2016).

5. Lustig-Yaeger, J. *et al.* A JWST transmission spectrum of the nearby Earth-sized exoplanet LHS 475 b. *Nat. Astron.* **7**, 1317–1328 (2023).

6. Zieba, S. *et al.* No thick carbon dioxide atmosphere on the rocky exoplanet TRAPPIST-1 c. *Nature* **620**, 746–749 (2023).

7. Fischer, D. A. *et al.* Five planets orbiting 55 Cancri. *Astrophys. J.* **675**, 790 (2008).

8. Dawson, R. I. & Fabrycky, D. C. Radial velocity planets de-aliased: a new, short period for super-Earth 55 Cnc e. *Astrophys. J.* **722**, 937 (2010).

9. Winn, J. N. *et al.* A super-Earth transiting a naked-eye star. *Astrophys. J. Lett.* **737**, 18 (2011).

10. Demory, B. O. *et al.* Detection of a transit of the super-Earth 55 Cancri e with warm Spitzer. *Astron. Astrophys.* **533**, 114 (2011).

11. Bourrier, V. *et al.* The 55 Cancri system reassessed. *Astron. Astrophys.* **619**, 1 (2018).

12. Crida, A., Ligi, R., Dorn, C., Borsa, F. & Lebreton, Y. Mass, radius, and composition of the transiting planet 55 Cnc e: using interferometry and correlations—A quick update. (2018).

13. Dorn, C., Harrison, J. H., Bonsor, A. & Hands, T. O. A new class of Super-Earths formed from high-temperature condensates: HD219134 b, 55 Cnc e, WASP-47 e. *Mon. Not. R. Astron. Soc.* **484**, 712–727 (2019).

14. Ehrenreich, D. *et al.* Hint of a transiting extended atmosphere on 55 Cancri b. *Astron. Astrophys.* **547**, 18 (2012).

15. Ridden-Harper, A. R. *et al.* Search for an exosphere in sodium and calcium in the transmission spectrum of exoplanet 55 Cancri e. *Astron. Astrophys.* **593**, 129 (2016).

16. Esteves, L. J., Mooij, E. J., Jayawardhana, R., Watson, C. & Kok, R. A search for water in a super-earth atmosphere: high-resolution optical spectroscopy of 55 Cancri e. *Astron. J.* **153**, 268 (2017).

17. Jindal, A. *et al.* Characterization of the Atmosphere of Super-Earth 55 Cancri e Using High-resolution Ground-based Spectroscopy. *Astron. J.* **160**, 101 (2020).

18. Tabernero, H. M. *et al.* HORuS transmission spectroscopy of 55 Cnc e. *Mon. Not. R. Astron. Soc.* **498**, 4222–4229 (2020).

19. Deibert, E. K. *et al.* A Near-infrared Chemical Inventory of the Atmosphere of 55 Cancri e. *Astron. J.* **161**, 209 (2021).

20. Keles, E. *et al.* The PEPSI exoplanet transit survey (PETS) I: investigating the presence of a silicate atmosphere on the super-earth 55 Cnc e. *Mon. Not. R. Astron. Soc.* **513**, 1544–1556 (2022).

21. Zhang, M. *et al.* No escaping helium from 55 Cnc e. *Astron. J.* **161**, 181 (2021).

22. Rasmussen, K. C. *et al.* A Nondetection of Iron in the First High-resolution Emission Study of the Lava Planet 55 Cnc e. *Astron. J.* **166**, 155 (2023).

23. Demory, B. O. *et al.* A map of the large day–night temperature gradient of a super-Earth exoplanet. *Nature* **532**, 207–209 (2016).

24. Demory, B. O., Gillon, M., Madhusudhan, N. & Queloz, D. Variability in the super-Earth 55 Cnc e. *Mon. Not. R. Astron. Soc.* **455**, 2018–2027 (2016).

25. Tamburo, P., Mandell, A., Deming, D. & Garhart, E. Confirming variability in the secondary eclipse depth of the super-Earth 55 Cancri e. *Astron. J.* **155**, 221 (2018).

26. Mercier, S. J., Dang, L., Gass, A., Cowan, N. B. & Bell, T. J. Revisiting the Iconic Spitzer

Phase Curve of 55 Cancri e: Hotter Dayside, Cooler Nightside, and Smaller Phase Offset. *Astron. J.* **164**, 204 (2022).

27. Angelo, I. & Hu, R. A case for an atmosphere on super-Earth 55 Cancri e. *Astron. J.* **154**, 232 (2017).

28. Brandeker, A. An asynchronous rotation scenario for 55 Cancri e. *AASDivision Extreme Sol. Syst. Abstr.* **51**, 311–07 (2019).

29. Schaefer, L. & Fegley, B. Chemistry of silicate atmospheres of evaporating super-Earths. *Astrophys. J.* **703**, 113 (2009).

30. Miguel, Y., Kaltenegger, L., Fegley, B. & Schaefer, L. Compositions of Hot Super-earth Atmospheres: Exploring Kepler Candidates. *Astrophys. J.* **742**, (2011).

31. Ito, Y. *et al.* Theoretical emission spectra of atmospheres of hot rocky super-Earths. *Astrophys. J.* **801**, 144 (2015).

32. Zilinskas, M. *et al.* Observability of evaporating lava worlds. *Astron. Astrophys.* **661, A126**, (2022).

33. Schlawin, E. *et al.* JWST noise floor. I. Random error sources in JWST NIRCam time series. *Astron. J.* **160**, 231 (2020).

34. Lally, M. & Vanderburg, A. Reassessing the Evidence for Time Variability in the Atmosphere of the Exoplanet HAT-P-7 b. *Astron. J.* **163**, 181 (2022).

35. Bell, T. J. *et al.* A first look at the JWST MIRI/LRS phase curve of WASP-43b. (2023).

36. Essack, Z., Seager, S. & Pajusalu, M. Low-albedo surfaces of lava worlds. *Astrophys. J.* **898**, 160 (2020).

37. Kipping, D. & Jansen, T. Detection of the Occultation of 55 Cancri e with TESS. *Res. Notes AAS* **4**, 170 (2020).

38. Demory, B. O. *et al.* 55 Cancri e's occultation captured with CHEOPS. *Astron. Astrophys.* **669**, 64 (2023).

39. Meier Valdés, E. A., Morris, B. M., Wells, R. D., Schanche, N. & Demory, B. O. Weak

evidence for variable occultation depth of 55 Cnc e with TESS. *Astron. Astrophys.* **663**, 95 (2022).

40. Meier Valdés, E. A. *et al.* Investigating the visible phase-curve variability of 55 Cnc e. *Astron. Astrophys.* **677**, 112 (2023).

41. Kite, E. S., Bruce Fegley Jr., Schaefer, L. & Gaidos, E. ATMOSPHERE-INTERIOR EXCHANGE ON HOT, ROCKY EXOPLANETS. *Astrophys. J.* **828**, 80 (2016).

42. Hammond, M. & Pierrehumbert, R. T. Linking the climate and thermal phase curve of 55 Cancri e. *Astrophys. J.* **849**, 152 (2017).

43. Zilinskas, M., Miguel, Y., Lyu, Y. & Bax, M. Temperature inversions on hot super-Earths: the case of CN in nitrogen-rich atmospheres. *Mon. Not. R. Astron. Soc.* **500**, 2197–2208 (2021).

44. Gaillard, F. *et al.* Redox controls during magma ocean degassing. *Earth Planet. Sci. Lett.* **577**, 117255 (2022).

45. Meier, T. G., Bower, D. J., Lichtenberg, T., Hammond, M. & Tackley, P. J. Interior dynamics of super-Earth 55 Cancri e. *Astron. Astrophys.* **678**, 29 (2023).

46. Heng, K. The Transient Outgassed Atmosphere of 55 Cancri e. *Astrophys. J. Lett.* **956**, 20 (2023).

47. Piette, A. A. *et al.* Rocky planet or water world? Observability of low-density lava world atmospheres. (2023).

**Methods**

**Data Analysis – NIRCam**

**Eureka! – Reduction 1**

We reduced the NIRCam data using the JWST Science Calibration Pipeline (*jwst*[48]) and *Eureka!*[49], with a few custom steps that we describe here. Starting from the uncal.fits files, we ran the default steps of *jwst* up to the dark current subtraction step. 1/f noise is a major source of noise in NIRCam data, particularly in the case of bright objects like 55 Cnc[33]. This electronic noise causes correlations in the fast-read direction (i.e., along detector rows in NIRCam). After running the dark current step, and to mitigate the 1/f noise in our data, we performed a row-by-row subtraction at the group level using the median of the first 650 pixels in each row. These pixels are not illuminated by the star and therefore provide a rough estimate of some of the correlated noise introduced during readout. We then fitted the ramps and performed the remaining default *jwst* calibration steps.

We fed the calibrated files (calints.fits) into *Eureka!*'s stage 3. We trimmed out the lowest 4 rows, which are reference pixels, and we extracted columns in the range 850–1945. We then corrected the curvature of the spectral trace by vertically shifting each column by a whole number of pixels. We performed a column-by-column background subtraction using the average value of the pixels located at least 23 pixels away from the source, rejecting 7σ outliers along the temporal and spectral axes. We extracted the optimal spectrum using an aperture region with a half-width of 6 pixels and a spatial profile constructed using the median frames. We computed the white light curve by binning the data in the 3.940–4.990 μm range. To generate the spectroscopic light curves, we binned the spectra into 21 bins of equal width. The number of bins is a compromise between the spectral resolution and having bins that are wide enough to divide each channel by the white light curve (see next paragraph).

The light curves are dominated by a downward ramp and time-correlated noise. We attempted two different methods to derive the emission spectrum of 55 Cnc e. In the first method, we fitted the white and each spectroscopic light curve using a *batman*[50] eclipse model, with the eclipse depth as a free parameter, multiplied by an exponential ramp and a linear polynomial. The detrended white light curve is shown in Fig. 1. We also included a white noise multiplier to scale the error bars according to the scatter of each light curve. The following parameters were kept fixed during the fits: orbital period and transit time (according to Ref[21]); planet radius, orbital inclination, and scaled semi-major axis (according to Ref[11]); and eccentricity and argument of periastron (0° and 90°, respectively). We trimmed the first 20 minutes of data.

In the second method, we divided each spectroscopic light curve by the white light curve, and we integrated the flux during the predicted time of the eclipse to produce a relative emission spectrum. The correlated noise is fairly consistent between the white and the spectroscopic light curves. As a consequence, dividing by the white light curve significantly reduces the correlated noise. We boosted the error bars of each resulting light curve to match the root-mean-square (RMS) of the residuals relative to a linear fit. The light curves were trimmed such that the baselines before and after eclipse have equal durations, and therefore our measurements of the relative eclipse depths are insensitive to any linear trends in time.

**Eureka! – Reduction 2**

We conducted data reduction using the *Eureka!* pipeline[49], with a few modifications at the initial stages of reduction. In the initial phase of data calibration, a row-by-row subtraction technique was implemented to eliminate correlated 1/f noise present in the dataset. This involved determining the median value of background pixels in each row and subtracting this value from the entire corresponding row. The remainder of the first two stages of data calibration adhered to the *Eureka!* procedure.

During Stage 3, the median value of background pixels was subtracted, specifically from rows 5 to 23 and 46 to 64, effectively reducing noise in the light curves. The subsequent stage involved the extraction of rows 26 to 43 and columns 740 to 2040. In the fourth stage, this data was used by the *Eureka!* pipeline to generate white and spectroscopic light curves in the wavelength range of 3.9365 to 4.9265 µm. A set of 30 spectroscopic light curves were created, followed by a process to eliminate outliers outside the 4-sigma rolling median within each spectroscopic light curve.

An observed declining trend in the light curves over time was attributed to an increase in the focal plane array housing temperature throughout the observation period. At this point, a departure from the *Eureka!* pipeline occurred for the purpose of detrending against this housing temperature. To address this, a power law fit between the fluxes and the housing temperature time series was carried out jointly with the eclipse model fit to the white light curve. An inverse correlation between the white light curve flux and the focal plane array housing temperature was observed throughout the observation period. The fit to the housing temperature time series was also divided out of each spectroscopic light curve. This process effectively removed the declining trend across all light curves. The fitting of the eclipse depth was then completed for each channel using *batman* in conjunction with *emcee*, with the transit time and orbital period fixed to the values reported in Ref[21], and with the planet radius, inclination, and scaled semi-major axis fixed to the values of Ref[11]. We also included a normalization term to maintain out-of-eclipse levels at 1 and a scaling factor multiplied into the uncertainties to accurately represent any correlated noise remaining in the spectroscopic light curves, bringing the total number of fitted parameters to 3.

**STARK**

We started our processing with the raw 'uncal' data and analyzed them using jwst pipeline with some changes such as omitting 'dark current' and 'flat field' steps because including them increased scatter in the final time series data. Since the data only have two groups per integration, the 'jump' step was redundant. We also skipped the 'photom' step as we are only interested in relative photometry. We performed background subtraction along columns and rows to reduce 1/f noise. Along the column, we fitted a line to the background pixels to estimate the background while we simply subtracted the median of background pixels along the rows. We manually searched for cosmic rays in the dataset by comparing the median frame with individual frames and replaced them with the average of neighboring pixels. All such pixels, along with those containing NaN and zeros, were added to the default bad pixel map produced by the jwst pipeline and ignored in further steps. We used the center-of-flux method to find the trace position in the data.

We used an open-source package *stark* (Spectral exTraction And Reduction Kit) to extract spectrum from the corrected data. *stark* robustly estimates the stellar point-spread function (PSF) by fitting splines to the data and then uses it to extract the spectrum. We fitted 1D and 2D splines as a function of distance from the trace and wavelength to the time series data to estimate the stellar PSF which was then used to estimate the spectra. Spectral apertures with a half width of 9 pixels and 2 pixels were used to compute PSF and extract the spectra, respectively. We repeated this whole procedure from background subtraction to spectral extraction iteratively. At the end of each iteration, the median static noise for each pixel (defined as the median difference over time between the data and synthetic frame estimated by the best-fit PSF and spectra) was subtracted from the raw data. We found that only two iterations were sufficient to robustly estimate the time series spectra.

We then computed the white-light lightcurve by calculating a weighted average of all spectral channels between 3.8612 and 4.9771 micron. We used *juliet*[51] to fit the lightcurve, which uses *batman*[50] and *dynesty*[52] to model the eclipse signal and to sample posteriors, respectively. In addition to the eclipse signal, we added linear, quadratic, and cubic polynomials in time to take into account long-term temporal trends in the data. All planetary parameters except for eclipse depth and eclipse time were fixed according to their values from the literature[11,39]. While we used a uniform prior between -500 and 500 ppm on eclipse depth, we put a normal prior on eclipse time based on Ref[39]. The eccentricity of the orbit was fixed to zero. We found the value of the best-fit eclipse depth to be 123±7 ppm. We then generated spectroscopic lightcurves by binning 23 columns (~22 nm), creating 50 spectroscopic channels. While fitting spectroscopic lightcurves, we fixed the eclipse time to the best-fit value from white-light lightcurve analysis. Except for this, the analysis was similar to the white-light lightcurve analysis.

In addition to this, we also computed relative eclipse depth spectra for a one-to-one comparison with spectra from different methods. We first generated 30 spectroscopic light curves using the same binning scheme as Eureka! - Reduction 2. We boosted error bars on the white-light light curve and spectroscopic light curves according to the light curve scatter, followed by dividing spectroscopic light curves with the white-light light curve. We finally took the ratio between the mean flux inside and outside of the eclipse duration and subtracted it from 1 to find relative eclipse depths. We made sure that the baseline on both sides of the eclipse signal is equal.

**SPARTA**

SPARTA is an independent pipeline that does not use any code from any other pipeline. It was originally developed to analyze MIRI/LRS data, and described in Ref[53]. We adapted SPARTA to analyze the NIRCam data obtained for 55 Cnc e. In the first stage, we subtracted the superbias, subtracted the reference pixels, performed non-linearity correction, subtracted the dark, and fit for the up-the-ramp reads (which, in this case, amounted to subtracting the first read from the second). In the second stage, we removed the background. We first attempted to remove 1/f noise to the maximum extent possible by subtracting the median of columns 4-600 of each row, from the entire row. We then removed any wavelength-dependent background by subtracting the median of the background region of each column, from the entire column. The background region was defined to be rows 4-11 (the first 4 are reference pixels) and 57-64, i.e., the 7 pixels closest to the top and bottom edges. In the third stage, we performed sum extraction. We tried the covariance-weighted optimal extraction algorithm[33] to mitigate 1/f noise, but did not see any substantial reduction in correlated noise. Unlike in Ref[53], where we summed the pixels within a predefined rectangular window, in this reduction we identified the trace and defined the window based on it. For each column, we fit a Gaussian to the spatial profile to identify the location of the trace; we then fit a second-degree polynomial to the trace location as a function of column number. We used this polynomial to define extraction windows with a half-width of 6 pixels, and took into account partial pixels when summing the flux within the window. In the fifth stage, we gathered all the spectra, computed normalized spectroscopic light curves, and masked >4-sigma outliers in each spectroscopic light curve.

The resulting white light curve exhibits high variability on ~hour timescales (Fig. 1). To obtain the white eclipse depth, we trimmed the first 30 minutes and fit the light curve with an exponential ramp (the amplitude and timescale both being free parameters), a linear trend, a linear function of the trace position in both directions, and an eclipse model computed by *batman* (with the eclipse depth being a free parameter[50]). The model fits poorly, indicating that the eclipse depth we obtain should be treated with caution. To obtain a relative eclipse spectrum, we divided the spectroscopic light curves by the white light curve, which, in addition to the higher noise in the spectroscopic light curves, made the correlated noise unnoticeable (Extended Data Fig. 1). We then trimmed the spectroscopic light curves until there was exactly as much data before the eclipse as after it. The relative eclipse depth was computed as $1-F_{in}/F_{out}$, where $F_{in}$ is the mean flux within the eclipse and $F_{out}$ is the mean flux out of the eclipse. Because we used the same out-of-eclipse baseline before and after the eclipse, any linear trend was automatically taken out.

We ensured that the final uncertainties of the relative eclipse depths adequately describe the uncertainties and scatters in the spectroscopic light curves. Based on the unbinned relative spectroscopic light curves, we calculated the RMS of the residual (fitting to a linear trend) and compared it with the mean uncertainty for each spectral channel. We found that a white error enhancement factor of 1.1 – 1.5 is necessary, and boosted the uncertainty of each data point by this factor. We then compared the RMS of the residual and the mean uncertainty for increasing binned sizes to the timescale of the eclipse (Extended Data Fig. 1). We found that they agree with each other very well, and concluded that the correlated noise is indeed unnoticeable in the relative spectroscopic light curves, after applying the white error enhancement factors. We then propagated the factors into the final relative depth uncertainties.

**Data Analysis – MIRI**

**Eureka!**

We reduced the MIRI data using *Eureka!*, with a modification to the code that we describe below. Starting from the uncalibrated raw data files, we ran stages 1 and 2 of *Eureka!*, skipping the jump detection step but running all other steps as default. Pixels values above a predefined saturation threshold are flagged by the pipeline and excluded from the ramp fits. After these calibration stages, we trimmed out rows and columns outside the 139–370 and 12–61 ranges, respectively. We masked pixels that were flagged as bad during stage 1, and we then measured the spatial position of the spectral trace on the detector in each integration with a Gaussian fit. We converted the data to electrons/second using a gain of 3.1 electrons/data number[53], instead of the default value of 5.5. We subtracted the background from each row using the mean value of pixels located at least 15 columns away from the source, excluding 5σ outliers along the time and spectral axes. We applied the outlier rejection technique to the full frame rather than just the background region.

We performed optimal extraction using an aperture region with a half-width of 5 pixels and a 60σ threshold for outlier rejection, and we used the non-smoothed median image as the spatial profile. After visually inspecting the data, we masked 2 bad pixel columns. We binned the data between 6.278 and 11.759 µm to extract the white light curve, and we produced 9 spectroscopic light curves in this range, with each spanning Δλ=0.609 µm. Below 6.278 µm, we could not produce robust measurements of the eclipse depths due to saturation, so these wavelengths were discarded from our analysis. We sigma clipped outliers in the light curves in the same manner as with the Eureka! Reduction 1 of the NIRCam data. The default routines of *Eureka!* were unable to robustly measure the drift in the spectral direction. To measure this value, relevant in the light curve fitting stage, we scaled and shifted the median frame to match each integration, as in SPARTA.

The MIRI light curves are strongly affected by systematics that mask the eclipse. There is a notable "V"-shaped drift in the trace position along the spectral and spatial directions whose duration and timing almost exactly match that of the occultation (Extended Data Fig. 2). The contribution of this drift to the light curves acts in the opposite direction as the eclipse of 55 Cnc e, hiding the drop in flux due to the planet disappearing out of view. Additionally, the light curves show a steep downward ramp. The exception is the two reddest channels, which are instead dominated by an upward ramp during the first ~20 minutes, characteristic of the "shadowed region" effect[35], followed by a more gradual downward ramp.

After trimming the first 30 minutes of data from each light curve to remove the integrations most affected by the ramps, we modeled each light curve using a combination of the *batman* eclipse model[50], a systematics model, and a white noise scaling factor. We adopted the same priors on the eclipse model parameters as in the light curve fits in the Eureka! - Reduction 1 of the NIRCam data, with the exception that in this case we did not fix the orbital period and transit time in the white light curve fit. Instead, we adopted Gaussian priors for these parameters based on the values in Ref[21]. In the spectroscopic light curve fits, we fixed these parameters to the values derived from the white light curve fit. We tested light curve fittings with and without a planetary phase variation component (modeled by a cosine peaking near eclipse and a sine peaking near quadrature, with their amplitudes as free parameters), and found that the fits without the phase component resulted in a larger Bayesian information criterion (BIC). We thus did not include the phase component in the final analysis.

Our systematics model included an exponential ramp, a linear polynomial, and a linear decorrelation against the drift in the spectral and spatial directions. All systematics parameters were fitted independently to each spectroscopic light curve and to the white light curve. To ensure that the linear decorrelation was not biased by the correlation of the eclipse with the "V"-shaped drift of the trace, we also estimated the coefficients after masking the data points taken during eclipse and found that they were consistent with the values derived from the full dataset (Extended Data Fig. 2).

The white and spectroscopic light curve residuals indicated the presence of time-correlated ("red") noise. To account for this red noise, we fitted the spectroscopic light curves again, this time fixing the white noise multiplier to the values derived in the first fit and inflating the uncertainties by a red noise multiplier. To calculate the red noise multiplier of each light curve, we computed the RMS of the binned residuals, with bin sizes ranging from 1 (i.e. no binning) to 6000 data points (approximately the eclipse duration). We divided the rms values by those expected in the absence of correlated noise[54], and we smoothed them with a 5-minute rolling-average filter to decrease noise. The maximum of the resulting values was taken as the red noise multiplier that we used to boost the error bars of each light curve, resulting in a conservative estimate of the contribution of the time-correlated noise to the final uncertainties. The red noise multiplier was 2.16 for the white light curve and lay between 1.02 and 1.79 for the spectroscopic light curves.

**SPARTA**

We made several changes from the methodology described in Ref[53]. Due to the brightness of 55 Cnc, we did not automatically exclude any groups from the beginning of the ramp, but did exclude groups that were above 50,000 DN (which we considered saturated). The 50,000 DN criterion was applied to the median uncalibrated integration, not to each individual integration, so that we always use the same number of groups for any given pixel. Another change we made to the up-the-ramp fitting is that instead of subtracting the median residuals of the fit for each segment (to "straighten out" the non-linearity of the ramp), we subtracted the median residuals of the fit for the entire dataset, which avoids the possibility of per-segment offsets. Finally, while Ref[53] used sum extraction, we used optimal extraction. The profile for optimal extraction was derived by computing the median image across all integrations and dividing each column by its sum. The half-width of the window used for optimal extraction was 5 pixels.

The MIRI emission spectrum was computed for wavelength bins linearly spaced in wavelength. The blue edge of the bluest bin was 6.278 µm, while the red edge of the reddest bin was 11.759 µm. The bins were defined so that the shadowed region, from 10.541 to 11.759 µm, is spanned by exactly two bins. Wavelength bins shortward of 6.278 µm were discarded because they include pixels which saturate in the second group, making it impossible to estimate their flux using the slope of the up-the-ramp samples. To compute the eclipse depth for a given wavelength bin, we fit the spectroscopic light curve with an exponential ramp (the amplitude and timescale being free parameters), a linear trend with time, a linear trend with the positions of the trace in the spatial and spectral directions, and an eclipse model computed by *batman*. We also estimated the red noise multiplier for each spectroscopic light curve using the same procedure as the Eureka! analysis, and inflated the error bars of the light curves by the red noise multiplier. The red noise multiplier lay between 1.08 and 2.40 for the spectroscopic light curves.

**Stellar Spectra**

We used the MIRI observation to derive the stellar spectrum in 5-12 µm. We used the zeroth group reads for wavelength <8 µm and all group reads for wavelength >8 µm to avoid any impact of saturation. The resulting stellar spectrum agrees with the previously published absolutely calibrated spectrum[55] in 5-6 µm within 10%. Ref[55] derived the spectrum from IRTF/SpeX observations with calibrations using the WISE W1 and W2 photometry in ~3-5 µm. We also verified that the MIRI measured stellar spectrum at 11.8 µm is consistent with the WISE W3 photometry[56] within 5%, and the spectrum is also consistent with the BT-Settl stellar model adopted by Ref[55] within 10% between 7-12 µm.

We also used the NIRCam observation to derive a stellar spectrum in 3.8-5 µm. We found that while the shape of the spectrum is reasonable, the flux at 5 µm measured by NIRCam is higher than that measured by MIRI by ~60%. This discontinuity may be due to imperfect calibration of the very bright star in the NIRCam measurement or in the saturated part of the MIRI measurement (5-6 µm). This uncertainty in the stellar flux at wavelengths <6 µm however does not impact our model interpretation because we allowed an arbitrary offset in the eclipse depth measured by NIRCam in both retrievals and comparison with self-consistent models. We notionally adopt the stellar spectrum of Ref[55] for wavelengths <5 µm and that measured by MIRI for wavelengths >5 µm for converting between the planetary thermal emission flux and the secondary eclipse depth in retrievals and self-consistent models.

**Retrievals**

We used a modified version of the open-source Python package PLATON[57] to assess the significance of atmospheric detections and derive atmospheric constraints from the measured thermal emission spectrum. We used the SPARTA analysis of the NIRCam observation and *Eureka!* analysis of the MIRI observation as the default for interpretation, and tested the sensitivity of the retrieval results by adopting the Eureka! – Reduction 1 analysis for the NIRCam observation. We allowed the absolute eclipse depth to vary as a free parameter while retaining the shape of the relative eclipse depth spectrum in the NIRCam bandpass.

We upgraded PLATON to retrieve the relative abundances of an arbitrary mixture of gases. Specifically, we sampled the abundance of gases in the centered log ratio (CLR) space with priors appropriate for the total number of considered gases[58] in each retrieval. We converted the CLRs to the volume mixing ratios (VMRs) for interpreting the physical results of our retrievals. We also used a simplified version of the analytical pressure-temperature profile of Ref[59] and it assumes one downwelling visible channel and uses a total of 3 parameters ($\beta$, $\log\kappa_{th}$, and $\log\gamma$), where $\beta$ represents the combined effect of the atmosphere's albedo, emissivity, and day-night redistribution (the same $\beta$ in Eq. 15 of Ref[59]), $\log\kappa_{th}$ is the log of the infrared opacity of the atmosphere, and $\log\gamma$ is the log of the ratio between the visible and infrared opacities. We did not assume an internal heat flux as an independent parameter for the temperature profiles for this highly irradiated planet (i.e., $T_{int}=0$). This simple version allows us to adequately model the pressure-temperature profiles given the quality of our data. The spectra are calculated at a resolution of R=1000 and then binned to match with the data from the JWST observations. To estimate the posteriors and Bayes evidence, PLATON uses the package *dynesty* to perform static nested sampling. We computed the Bayes factor using the Bayes evidence to determine the degree one scenario is preferred to another in terms of σ-confidence[60]. We used 5000 live points in the retrievals.

To ensure that the results of the spectral retrievals do not depend on the specific parametrization of the pressure-temperature profile, we additionally implemented a completely independent parameterization, where the surface temperature and the atmospheric temperature (assumed to be isothermal) determine the emission spectra[46]. The two-parameter model is probably the simplest temperature parameterization that one can design for retrieving emission spectra, and thus provides a good sanity check on our approach.

The null hypothesis is a blackbody planetary spectrum. To create a pure blackbody scenario within PLATON, we set the surface pressure to a minuscule $10^{-2}$ Pa with an isothermal atmosphere entirely composed of $N_2$ gas. At this pressure, $N_2$ is spectroscopically inactive in the infrared. This retrieval had two free parameters: (1) the blackbody temperature and (2) the NIRCam mean eclipse depth.

We then tried a range of plausible atmospheric compositions in retrievals. The common free parameters for each atmospheric retrieval were the 3 pressure-temperature profile parameters ($\beta$, $\log\kappa_{th}$, and $\log\gamma$), the surface pressure, and the NIRCam mean eclipse depth. We considered five single gas scenarios: $O_2$, $N_2$, $CO$, $CO_2$, and $SO_2$. The Bayes factor of each of these retrievals compared to the blackbody retrieval indicated a >3σ preference for the $CO_2$- and CO-only retrievals, with the $CO_2$-only scenario preferred at 3.7σ (Extended Data Fig. 4). The preference is slightly reduced when excluding the two red-most binned MIRI data points (corresponding to the shadow region), but remains above 3σ. The surface pressure in the CO scenario is larger than that in the $CO_2$ scenario (Extended Data Fig. 4), because a greater column of CO than $CO_2$ is needed to fit the spectral modulation in 4-5 μm. Using the alternative temperature parameterization resulted in the same preference for the $CO_2$ and CO scenarios (Extended Data Fig. 6).

Since atmospheres are not likely to be composed of one gas entirely, we also ran 14 two-gas atmosphere retrievals for $H_2O$, $CO_2$, SiO, $SO_2$, and CO with either $N_2$, CO, or $O_2$ as the background gas. We chose $O_2$ or $N_2$ as a background gas for their spectroscopic inactivity (while their collision-induced opacities become important at pressures >1 bar), and CO as a background gas because it is expected to be the dominant gas for an atmosphere outgassed from a magma ocean over a wide range of redox conditions[44]. Of these scenarios, the scenarios that include $CO_2$ (regardless of background gas) and the scenarios that use CO as the background gas are found to be >3σ preferred to the null hypothesis (Extended Data Fig. 4). The $CO_2$ + background gas scenarios are also >3σ preferred to the null hypothesis using the alternative temperature parameterization.

We thus focus on the $CO_2$ + background gas scenarios. Regardless of the background gas considered, the posteriors for the $CO_2$ runs are all generally consistent with each other. When comparing the $CO_2$ + $N_2$ and $CO_2$ + CO retrieval posteriors, we see that there is an extended tail towards lower $CO_2$ abundances when the background gas is CO (Extended Data Fig. 5). This is because both CO and $CO_2$ can induce absorption in the spectrum between 4 and 5 μm, and thus the amount of CO can increase to establish fits of similar goodness as the fits with more $CO_2$ and vice versa, while the wavelength position of the maximum absorption prefers $CO_2$ as the primary feature-creating gas.

When fitting only the NIRCam data, we found that all gas combinations tested can provide a substantially better fit than the blackbody model (Extended Data Fig. 6). Among these, the $N_2$-only and $O_2$-only scenarios fit the spectral modulation between 4-5 μm with $N_2$-$N_2$ or $O_2$-$O_2$ collision-induced opacities, which results in a very low absolute eclipse depth (<20 ppm). We thus deemed these scenarios unlikely. Both $SO_2$ and SiO have absorption bands between 4-5 μm and their existence can help improve the fit to the spectral modulation; however, they also have strong absorption bands in 5-12 μm and are thus constrained when the MIRI data are taken into account. The MIRI data are consistent with a blackbody model (Extended Data Fig. 6), while their absolute depths indicate a low temperature and thus efficient heat redistribution or multiple overlapping absorbers in the atmosphere. Therefore, both the spectral modulation from NIRCam and the eclipse depths from MIRI support a volatile atmosphere, and when put together, they prefer a $CO/CO_2$-rich one.

Even though $\log\kappa_{th}$ and $\log\gamma$ are taken as free parameters to drive pressure-temperature profiles in our retrievals, it is interesting to check whether their values are consistent with the retrieved gas abundances. Using a VMR of $10^{-3}$ for $CO_2$ at $10^4$ Pa and 2000 K, we calculated the mean opacity weighted by a Planck function of 2000 K and obtained a guiding value of $10^{-2}$ $m^2$ $kg^{-1}$ for $\kappa_{th}$. Using the mean opacity defined by Ref[61] results in a value of $10^{-4}$ $m^2$ $kg^{-1}$. These estimates are remarkably consistent with the retrieved posteriors (Extended Data Fig. 5). We also calculated the mean opacity weighted by a Planck function of 5200 K and obtained γ~0.1. This estimate does not include Rayleigh scattering, which would further increase γ. This check indicates that the pressure-temperature profiles used by the retrieval are physically plausible and also explains the generally consistent results between the retrievals and the grid search using self-consistent models.

Lastly, to test the sensitivity of the retrieval results on data reductions, we repeated a subset of retrieval runs using the Eureka! – Reduction 1 analysis for the NIRCam observation, as it has comparable final uncertainties as the SPARTA analysis. When using the NIRCam data only, the single-gas scenarios of $CO_2$ and CO and the binary mixture of $CO_2$ and CO are preferred to the blackbody null hypothesis by >2.9σ, and the most preferred scenario is the $CO_2$-only scenario at 3.1σ. When using both the NIRCam and the MIRI datasets, we also found that the single-gas scenario of $CO_2$ and the binary mixture of $CO_2$ and CO are preferred to the null hypothesis, with the single-gas scenario of $CO_2$ preferred by 2.6σ. These behaviors and overall trends are consistent with the findings from the SPARTA analysis for NIRCam, with a moderately lower degree of statistical preference for $CO_2$ and CO scenarios.

**Self-Consistent Models: Vaporized-Rock Atmosphere – Model 1**

At the temperature of 55 Cnc e, even if the planet has lost its entire inventory of highly volatile elements (H, C, N, S, P), the surface magma will outgas its relatively volatile components (e.g., Na, SiO, $SiO_2$, MgO) and eventually form a silicate atmosphere[29–32]. The initial step in the calculation of vaporized atmospheres involves determining the gases in equilibrium with the melt. With surface temperature, surface pressure, and melt composition as input, this equilibrium was computed using the LavAtmos code[62], which employs the MELTS code[63,64] to calculate the liquid oxide properties of the endmember species included in the melt ($SiO_2$, $Al_2O_3$, $TiO_2$, $Fe_2O_3$, $Fe_2SiO_4$, $Mg_2SiO_4$, $CaSiO_3$, $Na_2SiO_3$, and $KAlSiO_4$). While the Bulk Silicate Earth (BSE[65]) serves as our primary composition, we also examined other silicate-based melt compositions such as continental crust[66] and Mercury[67].

With elemental abundances established from the outgassing and equilibrium with the magma ocean, atmospheric chemistry was calculated using the FASTCHEM thermochemical equilibrium code[68]. This step operates in sync with the atmosphere's thermal structure. The selection of components is primarily constrained by available thermodynamic data, sourced from the Burcat NASA thermodynamics polynomial database.

We determined the temperature structure using the radiative transfer software HELIOS[69,70]. Through a self-consistent calculation with the chemistry, the profiles are adjusted to achieve a radiative-convective equilibrium. We assumed that the surface is non-reflective and that the surface temperature is influenced by the layers of the atmosphere above it. We considered 15 opacity sources (Al, AlO, Ca, CaO, Fe, K, Mg, MgO, Na, $O_2$, Si, SiO, $SiO_2$, Ti, TiO). Atomic opacities were obtained from the DACE database[71], utilizing VALD[72] and Kurucz[73] line lists. Molecular opacities were sourced from the DACE database or computed with HELIOS-K, fitted with Voigt profiles[71]. The specific line lists of all the species can be found in Ref[74].

Upon achieving a consistent solution for the thermal profile, chemistry, and outgassing, we modeled the synthetic emission spectra using the radiative transfer code petitRADTRANS[75]. The spectra were calculated at a resolution of R=λ/Δλ=1000 for the relevant wavelength region that encompasses the NIRCam and MIRI data. We used a pseudo-2D approach where the thermal emission of the planet is divided up into concentric rings of 10° wide, going from 0° to 80° zenith angle[76]. The spectrum of each of the rings is weighted according to the ring area and zenith angle and they are added up to produce the final spectrum.

None of silicate-based melt compositions considered can produce a thermal emission spectrum that fits the data (Extended Data Fig. 7). The spectra resulting from vastly different melt compositions are similar, but their flux levels significantly exceed the observed level for 55 Cnc e in the MIRI/LRS band. Also, the spectra do not show the spectral modulation in the NIRCam band. This comparison strongly suggests that the planet does not possess a thin, vaporized-rock atmosphere, but instead a more substantial atmosphere of volatile composition.

**Self-Consistent Models: Vaporized-Rock Atmosphere – Model 2**

We used the model of Ref[31] to simulate a vertically one-dimensional atmospheric structure with the vapor pressure of magma in the radiative, hydrostatic, and chemical equilibrium for given volatile-free (i.e., no highly volatile elements such as H, C, N, S, or Cl) magma composition, irradiation flux, and planet's properties. The MELTS code[63,64] was used to determine the chemical equilibrium condition between gases and magma at the ground, and the NASA CEA code[77] was used to compute the chemical equilibrium condition between gases in the atmosphere. The elemental abundances were assumed to be vertically constant. To obtain a pressure-temperature profile in radiative equilibrium, the two-stream equation with the assumption of quasi-isotropic radiation, adopting the δ−Eddington approximation[78], was integrated using the absorption opacity of Na, K, Fe, O, $O_2$, Si, and SiO. The opacity of $O_2$ was calculated using the absorption line data in the HITRAN database[79] and the opacity of other gases was calculated in the same way as Ref[80] using the Kurucz absorption line data[73]. To calculate the secondary eclipse spectra, we adopted the atmospheric structure at the cosine of zenith angle = 0.5 as the day-side averaged one. We refer the reader to Ref[31] for more information.

Extended Data Fig. 7 shows the simulated thermal emission spectrum of the rocky vapor atmosphere, assuming the Bulk Silicate Earth composition[81] for the magma composition and the irradiation flux and planet's properties of 55 Cnc e. The observed spectrum is not consistent with the vaporized rock atmosphere's spectrum, suggesting that the planet doesn't have such a vaporized rock atmosphere.

**Self-Consistent Models: Atmospheres with Varied CHONSP Abundances**

Ref[43] presented self-consistent models for atmospheres dominated by $N_2$, CO, and $H_2$ for 55 Cnc e. The models revealed that in C-H-N-O atmospheres with a high C/O ratio (>1) and low hydrogen content, temperature inversions could emerge due to shortwave absorbers such as CN. Increasing the hydrogen content, however, diminishes these inversions. Crucially, a higher carbon and oxygen proportion relative to $N_2$ can lead to pronounced $CO_2$ absorption at 4.3 μm, and this was also pointed out in a more recent study[47].

Building upon the model of Ref[43], we performed a parameter exploration of the abundances of C, H, O, N, S, and P elements to evaluate potential atmospheric compositions, pressure-temperature profiles, and thermal emission spectra. We modeled a wide range of compositions, starting from H-dominated cases and allowing the atmospheres to be increasingly C-, N- or O-dominant. We explored a mole fraction for C+O of $10^{-7}$, $10^{-3}$, 0.1, 0.5, 0.9, and ~1, with the C/O ratio of 0.1, 0.5, 1, and 10. In addition, we drastically varied S and P, allowing for up to 0.5 mole fraction for both. The minimum explored abundances for H and N were $10^{-8}$ and $10^{-5}$, respectively. Over 3500 self-consistent models were simulated to build the grid.

The chemical composition was calculated using the FastChem thermochemical equilibrium code[68], which includes condensation[82], resulting in atmospheres that are $H_2$-, CO-, $CO_2$-, $SO_2$-, and $N_2$-dominated. To determine the temperature structure, we used HELIOS[69,70] with the following volatile opacities: CO, $CO_2$, $O_2$, $H_2$, $N_2$, $C_2$, CH, $CH_3$, $CH_4$, $C_2H_2$, $C_2H_4$, CN, $H_2O$, $HNO_3$, $H_2O_2$, $H_2CO$, HCN, HS, $H_2S$, NO, $N_2O$, NH, $NH_3$, OH, S, CS, NS, SO, $SO_2$, $SO_3$, PH, $PH_3$, PS, PN, CP, and PO. The opacity also included the scattering of H, $H_2$, $H_2O$, $CO_2$, $N_2$, CO, $CH_4$, and $O_2$ and the continuum of $H^-$, as well as the collision-induced opacities of $H_2$-$H_2$, $H_2$-He, $CO_2$-$O_2$, $N_2$-$H_2$, $N_2$-$N_2$, and $O_2$-$O_2$ pairs. All of the opacities were obtained from the DACE[71], ExoMol[83], and HITRAN/HITEMP[84] databases, or were calculated using HELIOS-K[71]. The specific line lists used can be found in Ref[74]. The emission spectra were computed using petitRADTRANS[75], which employs the same opacity sources as described previously.

To model the planet's dayside emission, we used a heat redistribution factor f to encapsulate the impact of the heat transport. Applying the scaling law validated by general circulation models[61], we found that the heat redistribution factor f is close to the dayside limit of 2/3 when the surface pressure is 1 bar, and decreases to 0.4 for a surface pressure of 10 bar and 0.3 for a surface pressure of 100 bar. This behavior is consistent with the general circulation models specifically developed for 55 Cnc e[42]: with a high mean molecular weight atmosphere, the surface pressure needs to be >~10 bars to transport heat from the dayside to the nightside effectively. Meanwhile, with infrared absorbers in the atmosphere, the photosphere pressure for the thermal emission in 4-12 μm may well be smaller than the surface pressure by orders of magnitude. The effective heat redistribution factor then depends on complex interactions between horizontal and vertical transports and radiative transfer[85]. To explore a wide range of atmospheric composition, we assume a uniform surface pressure of 200 bars, and the heat redistribution factor of 2/3 and 0.3, covering the endmember scenarios from dayside to near-full redistribution.

Extended Data Fig. 8 shows the emission spectra of selected best-fit models and the corresponding pressure-temperature profiles. Models containing abundant $CO_2$, regardless of the background component, align well with the shape of the spectrum derived from the NIRCam observations. This is evident from the comparison of the $N_2$-$CO_2$-CO (red) and CO-$H_2$-$N_2$ (yellow) cases, as both reasonably fit the spectral modulation in 4-5 μm. The $CO_2$ mixing ratio in the CO-dominated case is more than two dex lower than that in the $N_2$-dominated case, because CO already provides absorption in 4-5 µm. In some of the fitting models, the $CO_2$ feature is complemented by the presence of $SO_2$ or a small amount of $H_2O$, allowing for a better fit to the spectrum in 4-4.2 µm. The best-fit models in the explored grid contain at least $10^{-5}$ $CO_2$ in volume mixing ratio, but the C/O ratio and the f factor are less constrained. We also find that models with a high $PH_3$ abundance (~$10^{-4}$ in volume mixing ratio) can sometimes result in a good overall fit.

Because the C-H-O-N-S-P atmospheres modeled here are likely connected to a magma ocean underneath, it is conceivable that rock-forming elements (Si, Mg, Fe, Al, Na, K, Ca, Ti) are vaporized from the magma ocean and mixed into the atmosphere. We modeled the impact of this phenomenon using the best-fit volatile-only model (red) as the baseline. We took two approaches to produce two scenarios shown in Extended Data Fig. 8 (black & purple spectra). The first of these (referred to as "Volatile + Outgassing") determined the atmospheric elemental abundances by summing vaporized-rock species and the assumed volatile composition[47,74]. The outgassed vaporized-rock species were calculated using the LavAtmos code[62]. For the second approach (referred to as "Inflated Silicates"), we inflated the amount of outgassing by coupling it to volatile species present in the atmosphere, resulting in a substantial increase in most outgassed species, particularly the short-wave absorbers (TiO, Na, and K). This leads to an inversion in the pressure-temperature profile of the atmosphere, transforming the absorption feature between 4 and 5 µm into an emission feature. While models with the inclusion of equilibrium outgassing can fit the NIRCam data, the increase in short-lived and sporadic outgassing of rock-forming elements from the magma ocean could explain the variability in eclipse depth measured by Spitzer in the 4.5 µm band.

Extended Data Fig. 8 also shows the minimum achieved $\chi^2$ values against the mole fraction of C+O and the C/O ratio. At the limit of inefficient heat redistribution (f=2/3), the models with large mole fractions of C+O (>~0.1) and sub-unity C/O ratios are preferred. These models have abundant $CO_2$ in the atmosphere. There is a singular outlying case of C+O=$10^{-7}$ for a mean NIRCam eclipse depth of 55 ppm. In this case the spectral modulation is fitted purely by $PH_3$ with a volume mixing ratio of $10^{-4}$, motivating future investigation into the possibility of the atmosphere being P-rich. With efficient heat redistribution (f=0.3), the models with C+O of 0.01-0.5 and sub-unity C/O ratios remain preferred, while the CO-dominated atmospheres (C/O>=1; C+O>=0.9) can also provide a reasonable fit. With efficient heat redistribution, the models with C/O>1 allow for graphite condensation, which lowers the gas-phase C/O ratio and keeps the atmosphere CO-dominated. These models typically have complex composition and pressure-temperature profiles, with subtle temperature variation in the photospheric region. For example, slight temperature inversions in the upper atmosphere can result in a more fine-tuned $CO_2$ feature, and together with strong CO absorption, providing a good fit to the observed NIRCam modulation.

In summary, the best-fit models center around three possible scenarios: a $CO_2$-rich atmosphere (C+O~$10^{-3}$, C/O<1), a CO-dominated atmosphere (C+O~1, C/O>=1), or a $PH_3$-rich atmosphere with minimal C+O influence (C/O=1, C+O=$10^{-7}$, $PH_3$~$10^{-4}$). The first scenario is uniquely favored when fitting the NIRCam data alone (Extended Data Fig. 8), which is consistent with spectral retrievals. By contrast, the MIRI data does not indicate any clear molecular features, suggesting either efficient heat redistribution or overlapping absorption features (e.g., $H_2O$ in 7-9 μm and $CO_2$ in 9-11 μm) that place the photosphere to the cooler regions of the atmosphere.

**Self-Consistent Models: Atmospheres in Equilibrium with Magma Oceans**

We modeled an atmosphere in equilibrium with the magma ocean, using a magma ocean-atmosphere partitioning and chemical equilibrium/speciation model[44]. The model simulates the C-H-N-S mass fraction in the atmosphere and that dissolved in the magma ocean, with the oxygen fugacity of the magma as a free parameter. $H_2O$, $H_2$, $O_2$, $CO_2$, CO, $CH_4$, $N_2$, $H_2S$, $S_2$, and $SO_2$ are considered in chemical equilibrium and their solubilities in the magma are applied.

We assumed the volatile content of H, C, N, and S as Earth's early magma ocean (a combination of surficial volatile inventory and the volatile content of the depleted mantle, following the 1/2 BSM case of Ref[44]), and the size of the magma ocean at 55 Cnc e to be 1/3 of the planet mass (i.e., 1/2 of the mass of the mantle if the planet has a core/mantle mass fraction similar to Earth, or a magma ocean depth of ~2000 km). The planet can have a smaller magma ocean of ~500 km (~9% of the planet's mass) if the temperature profile is sub-adiabatic[45]. We found that the resulting atmospheric size (i.e., the total pressure) and composition only weakly depend on the assumed magma ocean size. We explored an oxygen fugacity of the magma ranging from highly reduced (IW-6) to oxidized (IW+4), where IW stands for the iron-wustite buffer. Extended Data Fig. 9 shows that the atmosphere is CO-dominated for an oxygen fugacity between IW-4 and IW, $CO_2/SO_2$-dominated if more oxidized, and $H_2$-dominated if more reduced. The total atmospheric pressure ranges in 100 – 600 bars, consistent with the inference based on the planet's mass and radius[11,12]. The total atmospheric pressure is approximately proportional to the volatile content (e.g., 1% of the volatile content would result in an atmosphere of ~ 1 bar). When decreasing the volatile content, we found a more-than-proportional decrease in atmospheric $H_2$ (and also $H_2O$, $CH_4$, and $H_2S$), and a less-than-proportional decrease in atmospheric $SO_2$. This is consistent with the expectation from different solubilities of water, carbon, and sulfur as a function of pressure in the magma[86].

In another scenario, we considered that the planet has lost most of its H endowment, with only 3% H content of the standard model. This results in substantially less $H_2O$ and $H_2$ in the atmosphere, and practically removes $CH_4$ and $H_2S$, compared to the standard model (Extended Data Fig. 9). The partial pressures of CO, $CO_2$, $N_2$, and $SO_2$ do not change.

We then used the ExoPlanet Atmospheric Chemistry & Radiative Interaction Simulator (EPACRIS) to compute the pressure-temperature profile and molecular abundance profiles. EPACRIS was developed based on a gray-atmosphere and thermochemical equilibrium atmospheric model[87]. This model was updated with a new radiative-convective climate model, solving for pressure-temperature profiles. The radiative transfer scheme solves the 2-stream radiative fluxes formulated by Ref[88] with a pentadiagonal flux solver, PTRANS-I[89]. The radiative equilibrium is solved iteratively with the Newton-Raphson method. We checked for the Schwarzschild criterion to assess convective stability and performed adiabatic adjustment, conserving enthalpy through the use of potential temperatures for each convective region for each radiative-convective simulation[70]. We did not include any condensates in these models.

Given high temperatures, thermochemical equilibrium is sufficient to model the planet's atmosphere[87]. We used elemental abundances summarized from the magma ocean calculations (Extended Data Fig. 9) as input parameters for the atmospheric models. Since magma typically has very low albedo in the visible[36], we assumed the magma ocean surface to be completely dark (i.e., zero albedo for any starlight that reaches the surface and emissivity of unity in the thermal infrared). Tidal heating could provide additional energy to the bottom of the atmosphere and we include an internal heat flux of 100 K, guided by tidal heating models[90], which is much smaller than heating from stellar irradiation. For opacities, we included the molecular absorptions of $H_2O$, $CH_4$, $NH_3$, $CO_2$, $O_2$, OH, $H_2CO$, $H_2O_2$, CO, $O_3$, $C_2H_2$, $C_2H_4$, $C_2H_6$, HCN, $N_2O$, NO, $NO_2$, $HNO_3$, $H_2S$, $SO_2$, and OCS, as well as $H_2$-$H_2$, $H_2$-H, $N_2$-$N_2$, $N_2$-$H_2$, and $CO_2$-$CO_2$ collision-induced absorption, computed from HITRAN 2012 and HITEMP 2010 databases. The radiative-convective scheme and the chemical equilibrium scheme were applied iteratively to achieve converged solutions.

Extended Data Fig. 9 shows two sets of models, with one standard set assuming a 200-bar atmosphere and Earth-like C-H-N-S volatile abundance ratios in the atmosphere-magma ocean system, and another set assuming a 1-bar atmosphere and an H abundance ratio reduced to 3% of the standard cases. The second set is designed to mimic a scenario where the planet is depleted of C-H-N-S volatiles (to ~1% of bulk silicate Earth's volatile abundances) and H is additionally depleted because it is preferentially lost to space. All scenarios in the standard model set provide an adequate fit to the data, because all of them have considerable mixing ratios of CO and $CO_2$ in the atmosphere and can fit the spectral modulation observed at 4-5 μm. There is a slight preference for the more reducing scenarios, and this is because the scenarios more oxidizing than IW have abundant $SO_2$ in the atmosphere, which is disfavored by the data among this set of models. On the other hand, in the 1-bar, reduced-H model set, the atmosphere is so thin and $H_2O$/$CH_4$-poor that infrared windows to the surface appear at 3-4.2 μm and 6-9 μm. As a result, only the more oxidized models that have $SO_2$ can provide a good fit to the data. Regardless of the oxygen fugacity, the exchange with the magma ocean ensures that the atmosphere is rich in CO and has a $CO_2$ mixing ratio ranging from $10^{-1}$ to $10^{-4}$, and this is remarkably consistent with the scenarios preferred by both spectral retrievals and the atmospheric model grid with varied CHONSP abundances.

**An Atmosphere Sustained by Magma Ocean Outgassing**

Adopting an X-ray and extreme ultraviolet luminosity of $10^{27.7}$ erg s$^{-1}$ for 55 Cnc[21,91], and an energy-limited escape efficiency of 0.1[92], we estimated the escape rate from 55 Cnc e to be approximately $10^9$ g s$^{-1}$, comparable to hydrodynamic escape simulations[93]. The rapid escape results in a very short lifetime of approximately 2 million years for a 10-bar atmosphere. A recent study showed that only 0.1% of the stellar X-ray and extreme ultraviolet irradiation can be converted to hydrodynamic motion in an $N_2$-/$O_2$-rich atmosphere at an irradiation environment 1000 times more intense than on Earth (approximately corresponding to 55 Cnc e)[94]. If the very low efficiency is applied, the lifetime would increase to 200 million years, remaining small compared to the current estimated system age of 8 billion years[95].

The atmosphere of 55 Cnc e can however be sustained by magma ocean outgassing. Adopting an age dependency of $t^{-0.9}$ for the X-ray flux and $t^{-1.3}$ for the extreme ultraviolet flux, and assuming the escape rate to be a constant in the first 1 billion years, we found that 2.8% of the planetary mass would be lost due to escape within 8 billion years for the efficiency of 0.1. In comparison, primitive chondrite meteorites (CI and CM type) have 2-6% carbon content[96]. Earth's bulk carbon content is only ~0.01%[97], and this volatile content would already produce a ~200 bar CO-$CO_2$ atmosphere in equilibrium with the magma ocean. An evolutionary scenario where the planet lost a large fraction of its volatile endowment is thus consistent with our findings of a secondary volatile-rich atmosphere.

Furthermore, elemental fractionation of the outflow will help retain C, N, O, etc volatiles. Whether the escape of hydrogen can drag along heavier elements depends on how the overall escape rate compares with the diffusion-limited escape rate[98–100]. Using the escape rate for the efficiency of 0.1, we found that it is much larger than the diffusion-limited escape rate between H and He, implying that He escapes at nearly the same speed as H[101]. However, we found that it is only 20% larger than the diffusion-limited escape rate between H and O[98] (taken as a proxy for C, N, S, P), implying that the escape of these elements is only a fraction of that of H, scaled by their instantaneous mixing ratios. A higher escape rate in the past would reduce the fractionation, but a lower efficiency[94] would amplify the fractionation. It is therefore likely that H and He have been preferentially lost from 55 Cnc e, leaving behind enhanced C and N and oxidizing the atmosphere-magma ocean system.

The analysis here on 55 Cnc e, one of the most irradiated rocky planets, shows that it is hard to remove the complete inventory of volatiles from a planet. Rocky planets with a permanent magma ocean, as opposed to rocky planets with a solid surface, thus provide a promising avenue to detect outgassed atmospheres and a window to study atmosphere-mantle exchange.

**Limited Contribution of Potential Circumstellar Dust**

Visible-wavelength observations of 55 Cnc indicated variability[39,40,102,103], and a circumstellar torus of dust made of exotic material that can stay solid in the highly irradiated environment near 55 Cnc e[40] similar to Jupiter-Io system[104] is a possible scenario. The variability may also have a stellar origin due to, for example, supergranulation[34]. In the CHEOPS bandpass, the light curve phase-folded according to the orbital period of 55 Cnc e has a phase amplitude of ~15 ppm (variable up to 50 ppm) with no consistent secondary eclipse, suggesting that the origin of the signal is unlikely to be a static planetary atmosphere or surface[40]. At the temperature of the planet found in this work (~2000 K), the planet's thermal emission contribution to the signal in the CHEOPS bandpass is <3 ppm (and similarly for MOST or TESS).

Here we estimate the contribution of this dusty torus, if it exists, to the thermal emission measurements of JWST. Following the torus model in Section 4.4 of Ref[40], we let f be the magnitude of the phase modulation in the visible, and assume that the light-curve modulation comes from dust occulting the star. Then, the total number of dust particles is

$N_d = (R_*/R_d)^2 f$,

where $R_d$ is the mean dust particle size, and $R_*$ is the radius of the star. This formula is equivalent to Eq. (8) of Ref[40].

In reality, the dusty torus may be "non-transiting" and the visible-light-curve modulation may instead come from the reflection of the dust particles. In this case,

$f = N_d' (R_d/a)^2 A$,

where a is the semi-major axis, and A is the albedo of the dust particle. It is trivial to show that

$N_d' = N_d (a/R_*)^2 (1/A)$.

For 55 Cnc e, $(a/R_*)^2=12$, and thus the total number of dust particles would need to be 1-2 orders of magnitude larger than the occulting dust scenario.

Turning to the thermal emission, the magnitude of the light-curve modulation caused by the torus can be estimated as

$f_{thermal} = N_d (R_d/R_*)^2 B(T_d)/B(T_*) = f B(T_d)/B(T_*)$,

where B is the Planck function for the temperature of the particle ($T_d$) or the star ($T_*$). For $T_d$=2500 K and $T_*$=5200 K, this ratio of the Planck function is 0.3 at 4 μm and 0.4 at 10 μm. Therefore, if the visible-wavelength phase modulation is caused by a dusty torus occulting the star, the contribution of this dusty torus to the JWST observations is at most 5-10 ppm, which is small compared to the occultation depth observed in the MIRI band.

However, if the visible-wavelength phase modulation is caused by a more massive dusty torus reflecting the starlight, the contribution to the JWST observation could be 1-2 orders of magnitude larger, to >50 ppm. This would impact JWST observations. The required mass loss rate to support such a dusty torus would be $4\times10^{12} - 3\times10^{14}$ kg/year, or $10^5 - 10^7$ kg/s[40]. Tidal heating as a surface source can supply at least ~$10^6$ kg/s[90,104], which is technically in line with the requirement. The supply from the surface can be far larger due to the sublimation processes, while it is unclear what mechanism would lift the dust to space, especially when a volatile-rich atmosphere is present. Neglecting plasma processes, and assuming radiation pressure is the dominant loss mechanism, Ref[40] found that only select minerals (e.g., alpha quartz, silicon carbide) survive for parts of the orbit at 55 Cnc e. Although intriguing, we believe that this scenario requires a more detailed study to be conclusive.

Alternatively, the visible-light variability may be produced by a plasma torus confined by a stellar magnetic field (e.g., $CY^+$, $NaX^+$)[105]. This plasma torus is sustained by atmospheric or exospheric escape from the planet[106]. The charged particles may have visible-light opacities and drive the stochastic variation seen thus far[38,102]. Although ions do not generally have continuum opacities in the thermal infrared, exospheric neutrals may provide thermal emission opacities over a small bandpass.

**Limited Contribution of the Non-Transiting Planet b**

Assuming that for the Jupiter-mass planet b, $R_p \sim 1\ R_{Jupiter}$, then $R_p/R_* \sim 0.1$. Using its semi-major axis of a=0.1134, we estimate that $T_{eq}$=738 K for zero albedo and full thermal redistribution. The maximum contribution to the light curve in thermal infrared is $F_p/F_* < (R_p/R_*)^2 * B(T_{eq})/B(T_*)$=120 ppm at 4.5 µm and 380 ppm at 7.5 µm. For a period of 14.65 days, the change in the contributed flux during our 0.23-day observation is at most 1.9 ppm at 4.5 µm and 6.0 ppm at 7.5 µm. However, there could be short-time variability in the atmosphere of the planet b that manifests as variability of the light curve.

**Additional Reference in Methods**

48. Bushouse, H. *et al.* JWST Calibration Pipeline. (2022) doi:10.5281/ZENODO.7038885.

49. Bell, T. J. *et al.* Eureka!: an end-to-end pipeline for JWST time-series observations. (2022).

50. Kreidberg, L. batman: basic transit model calculation in Python. *Publ. Astron. Soc. Pac.* **127**, 1161 (2015).

51. Espinoza, N., Kossakowski, D. & Brahm, R. Juliet: a versatile modelling tool for transiting and non-transiting exoplanetary systems. *Mon. Not. R. Astron. Soc.* **490**, 2262–2283 (2019).

52. Speagle, J. S. DYNESTY: a dynamic nested sampling package for estimating Bayesian posteriors and evidences. *Mon. Not. R. Astron. Soc.* **493**, 3132–3158 (2020).

53. Kempton, E. M. R. *et al.* A reflective, metal-rich atmosphere for GJ 1214b from its JWST phase curve. *Nature* 1–2 (2023).

54. Winn, J. N. *et al.* The Transit Light Curve Project. IX. Evidence for a Smaller Radius of the Exoplanet XO‐3b. *Astrophys. J.* **683**, 1076–1084 (2008).

55. Crossfield, I. J. ACME stellar spectra-I. Absolutely calibrated, mostly empirical flux densities of 55 Cancri and its transiting planet 55 Cancri e. *Astron. Astrophys.* **545**, 97 (2012).

56. Cutri, R. M. *et al. Explanatory Supplement to the AllWISE Data Release Products*. *Explanatory Supplement to the AllWISE Data Release Products* 1 https://ui.adsabs.harvard.edu/abs/2013wise.rept....1C (2013).

57. Zhang, M., Chachan, Y., Kempton, E. M. R., Knutson, H. A. & Chang, W. PLATON II: New Capabilities and a Comprehensive Retrieval on GH 189733b Transit and Eclipse Data. *Astrophys. J.* **899**, 27 (2020).

58. Damiano, M. & Hu, R. Reflected Spectroscopy of Small Exoplanets I: Determining the Atmospheric Composition of Sub-Neptunes Planets. *Astron. J.* **162**, 200 (2021).

59. Line, M. R. *et al.* A Systematic Retrieval Analysis of Secondary Eclipse Spectra I. A Comparison of Atmospheric Retrieval Techniques. *Astrophys. J.* **775**, 137 (2013).

60. Benneke, B. & Seager, S. How to Distinguish between Cloudy Mini-Neptunes and Water/Volatile-dominated Super-Earths. *Astrophys. J.* **778**, 153 (2013).

61. Koll, D. D. A scaling for atmospheric heat redistribution on tidally locked rocky planets. *Astrophys. J.* **924**, 134 (2022).

62. Buchem, C. P. A., Miguel, Y., Zilinskas, M. & Westrenen, W. LavAtmos: An open-source chemical equilibrium vaporization code for lava worlds. *Meteorit. Planet. Sci.* **58**, 1149–1161 (2023).

63. Ghiorso, M. S. & Sack, R. O. Chemical Mass Transfer in Magmatic Processes. IV. A Revised and Internally Consistent Thermodynamic Model for the Interpolation and Extrapolation of Liquid-Solid Equilibria in Magmatic Systems at Elevated Temperatures and Pressures. *Contrib. Mineral. Petrol.* **119**, 197–212 (1995).

64. Asimow, P. D. & Ghiorso, M. S. Algorithmic Modifications Extending MELTS to Calculate Subsolidus Phase Relations. *Am. Mineral.* **83**, 1127–1131 (1998).

65. Palme, H. & O'Neill, H. S. C. Cosmochemical Estimates of Mantle Composition. in *Treatise on Geochemistry* (eds. Holland, H. D. & Turekian, K. K.) vol. 2 38 (Elsevier Ltd: Pergamon, 2003).

66. Wedepohl, K. H. Geochim. *Cosmochim Acta* **59**, 1217 (1995).

67. Morgan J. W. & Anders E. Proc. *Natl Acad Sci USA* **77**, 6973 (1980).

68. Stock, J. W., Kitzmann, D. & Patzer, A. B. C. FastChem 2: an improved computer program to determine the gas-phase chemical equilibrium composition for arbitrary element distributions. *Mon. Not. R. Astron. Soc.* **517**, 4070–4080 (2022).

69. Malik, M. *et al.* HELIOS: an open-source, GPU-accelerated radiative transfer code for self-consistent exoplanetary atmospheres. *Astron. J.* **153**, 56 (2017).

70. Malik, M. *et al.* Self-luminous and irradiated exoplanetary atmospheres explored with

HELIOS. *Astron. J.* **157**, 170 (2019).

71. Grimm, S. L. *et al.* HELIOS-K 2.0 opacity calculator and open-source opacity database for exoplanetary atmospheres. *Astrophys. J. Suppl. Ser.* **253**, 30 (2021).

72. Ryabchikova, T. *et al.* A major upgrade of the VALD database. *Phys. Scr.* **90**, 054005 (2015).

73. Kurucz, R. L. Atomic and molecular data for opacity calculations. *Rev. Mex. Astron. Astrofisica* **23**, 23 (1992).

74. Zilinskas, M., Miguel, Y., Buchem, C. P. A. & Snellen, I. A. G. Observability of silicates in volatile atmospheres of super-Earths and sub-Neptunes-Exploring the edge of the evaporation desert. *Astron. Astrophys.* **671**, 138 (2023).

75. Mollière, P. *et al.* petitRADTRANS-A Python radiative transfer package for exoplanet characterization and retrieval. *Astron. Astrophys.* **627**, 67 (2019).

76. Zieba, S. *et al.* K2 and Spitzer phase curves of the rocky ultra-short-period planet K2-141 b hint at a tenuous rock vapor atmosphere. *Astron. Astrophys.* **664**, 79 (2022).

77. Gordon, S. & McBride, B. J. *The NASA Computer Program CEA (Chemical Equilibrium with Applications*. (NASA Reference Publication, 1996).

78. Toon, O. B., McKay, C. P., Ackerman, T. P. & Santhanam, K. Rapid calculation of radiative heating rates and photodissociation rates in inhomogeneous multiple scattering atmospheres. *J. Geophys. Res. Atmospheres* **94**, 16287–16301 (1989).

79. Rothman, L. S. & Gordon, I. E. The HITRAN molecular database. *AIP Conf. Proc.* **1545**, 223–231 (2013).

80. Piskunov, N. & Kupka, F. Model atmospheres with individualized abundances. *Astrophys. J.* **547**, 1040 (2001).

81. McDonough, W. F. & Sun, S. S. The composition of the Earth. *Chem. Geol.* **120**, 223–253 (1995).

82. Kitzmann, D., Stock, J. W. & Patzer, A. B. C. Fastchem Cond: equilibrium chemistry with

condensation and rainout for cool planetary and stellar environments. *Mon. Not. R. Astron. Soc.* **527**, 7263–7283 (2023).

83. Chubb, K. L. *et al.* The ExoMolOP database: Cross sections and k-tables for molecules of interest in high-temperature exoplanet atmospheres. *Astron. Astrophys.* **646**, 21 (2021).

84. Gordon, I. E. *et al.* The HITRAN2016 molecular spectroscopic database. *J. Quant. Spectrosc. Radiat. Transf.* **203**, 3–69 (2017).

85. Perez-Becker, D. & Showman, A. P. ATMOSPHERIC HEAT REDISTRIBUTION ON HOT JUPITERS. *Astrophys. J.* **776**, 134 (2013).

86. Gaillard, F. & Scaillet, B. A theoretical framework for volcanic degassing chemistry in a comparative planetology perspective and implications for planetary atmospheres. *Earth Planet. Sci. Lett.* **403**, 307–316 (2014).

87. Hu, R. & Seager, S. Photochemistry in terrestrial exoplanet atmospheres. III. Photochemistry and thermochemistry in thick atmospheres on super Earths and mini Neptunes. *Astrophys. J.* **784**, 63 (2014).

88. Heng, K., Malik, M. & Kitzmann, D. Analytical models of exoplanetary atmospheres. VI. Full solutions for improved two-stream radiative transfer, including direct stellar beam. *Astrophys. J. Suppl. Ser.* **237**, 29 (2018).

89. Askar, S. S. & Karawia, A. A. On solving pentadiagonal linear systems via transformations. *Math. Probl. Eng.* (2015).

90. Bolmont, E. *et al.* Tidal dissipation and eccentricity pumping: Implications for the depth of the secondary eclipse of 55 Cancri e. *Astron. Astrophys.* **556**, 17 (2013).

91. Salz, M., Czesla, S., Schneider, P. C. & Schmitt, J. H. M. M. Simulating the escaping atmospheres of hot gas planets in the solar neighborhood. *Astron. Astrophys.* **586**, A75 (2016).

92. Kubyshkina, D. *et al.* Grid of upper atmosphere models for 1–40 $M_\oplus$ planets: application to CoRoT-7 b and HD 219134 b,c. *Astron. Astrophys.* **619**, A151 (2018).

93. Lammer, H. *et al.* Probing the blow-off criteria of hydrogen-rich 'super-Earths'. *Mon. Not. R. Astron. Soc.* **430**, 1247–1256 (2013).

94. Nakayama, A., Ikoma, M. & Terada, N. Survival of Terrestrial N2–O2 Atmospheres in Violent XUV Environments through Efficient Atomic Line Radiative Cooling. *Astrophys. J.* **937**, 72 (2022).

95. Mamajek, E. E. & Hillenbrand, L. A. Improved age estimation for solar-type dwarfs using activity-rotation diagnostics. *Astrophys. J.* **687**, 1264 (2008).

96. Pearson, V. K., Sephton, M. A., Franchi, I. A., Gibson, J. M. & Gilmour, I. Carbon and nitrogen in carbonaceous chondrites: Elemental abundances and stable isotopic compositions. *Meteorit. Planet. Sci.* **41**, 1899–1918 (2006).

97. Hirschmann, M. M. Comparative deep Earth volatile cycles: The case for C recycling from exosphere/mantle fractionation of major. *Earth Planet. Sci. Lett.* **502**, 262–273 (2018).

98. Zahnle, K. J. & Kasting, J. F. Mass fractionation during transonic escape and implications for loss of water from Mars and Venus. *Icarus* **68**, 462–480 (1986).

99. Hunten, D. M., Pepin, R. O. & Walker, J. C. G. Mass fractionation in hydrodynamic escape. *Icarus* **69**, 532–549 (1987).

100. Zahnle, K., Kasting, J. F. & Pollack, J. B. Mass fractionation of noble gases in diffusion-limited hydrodynamic hydrogen escape. *Icarus* **84**, 502–527 (1990).

101. Hu, R., Seager, S. & Yung, Y. L. HELIUM ATMOSPHERES ON WARM NEPTUNE- AND SUB-NEPTUNE-SIZED EXOPLANETS AND APPLICATIONS TO GJ 436b. *Astrophys. J.* **807**, 8 (2015).

102. Sulis, S. *et al.* Multi-season optical modulation phased with the orbit of the super-Earth 55 Cancri e. *Astron. Astrophys.* **631**, 129 (2019).

103. Morris, B. M. *et al.* CHEOPS precision phase curve of the Super-Earth 55 Cancri e. *Astron. Astrophys.* **653**, 173 (2021).

104. Oza, A. V. *et al.* Sodium and potassium signatures of volcanic satellites orbiting close-in

gas giant exoplanets. *Astrophys. J.* **885**, 168 (2019).

105. Folsom, C. P. *et al.* Circumstellar environment of 55 Cancri-The super-Earth 55 Cnc e as a primary target for star–planet interactions. *Astron. Astrophys.* **633**, 48 (2020).

106. Gebek, A. & Oza, A. V. Alkaline exospheres of exoplanet systems: evaporative transmission spectra. *Mon. Not. R. Astron. Soc.* **497**, 5271–5291 (2020).

**Data Availability**

The data used in this paper are associated with JWST guest observer program 1952 and are available from the Mikulski Archive for Space Telescopes (https://mast.stsci.edu). The data products required to generate Figs. 1-3 and Extended Data Figs. 1-9, as well as the stellar spectrum and the data reduction configuration files for the Eureka! - Reduction 1 and SPARTA analyses are available here: https://osf.io/2s6md/ with DOI:10.17605/OSF.IO/2S6MD. All additional data are available upon request.

**Code Availability**

The codes used in this publication to extract, reduce and analyze the data are as follows: STScI JWST calibration pipeline (https://github.com/spacetelescope/jwst), Eureka! (https://eurekadocs.readthedocs.io/en/latest/), STARK (https://github.com/Jayshil/stark), SPARTA (https://github.com/ideasrule/sparta), batman (http://lkreidberg.github.io/batman/docs/html/index.html), emcee (https://emcee.readthedocs.io/en/stable/), dynesty (https://dynesty.readthedocs.io/en/stable/index.html), and juliet (https://juliet.readthedocs.io/en/latest/). In addition, we have made use of HELIOS (https://github.com/exoclime/HELIOS), FastChem (https://github.com/exoclime/FastChem), PLATON (https://github.com/ideasrule/platon), petitRADTRANS (http://gitlab.com/mauricemolli/petitRADTRANS), and LavAtmos (https://github.com/cvbuchem/LavAtmos) to produce models.

**Inclusion & Ethics**

All authors have committed to upholding the principles of research ethics & inclusion as advocated by the Nature Portfolio journals.

**Acknowledgement**

We thank Fabrice Gaillard for helpful discussion on magma ocean outgassing and Julie Inglis for a helpful discussion regarding data reduction with JWST. This research is based on observations with the NASA/ESA/CSA James Webb Space Telescope (JWST) obtained from the Mikulski Archive for Space Telescopes at the Space Telescope Science Institute, which is operated by the Association of Universities for Research in Astronomy, Incorporated, under NASA contract NAS 5-03127. These observations are associated with program no. JWST-GO-1952. Support for program no. JWST-GO-1952 was provided through a grant from the STScI under NASA contract NAS 5-03127. Part of this research was carried out at the Jet Propulsion Laboratory, California Institute of Technology, under a contract with the National Aeronautics and Space Administration (80NM0018D0004). Part of the high performance computing resources used in this investigation were provided by funding from the JPL Information and Technology Solutions Directorate. M.Z. acknowledges support from the 51 Pegasi b Fellowship financed by the Heising-Simons Foundation. Y.M. and M.Z. have received funding from the European Research Council (ERC) under the European Union's Horizon 2020 research and innovation programme (grant agreement no. 101088557, N-GINE). Y.M. and C.B. acknowledge support of a Dutch Science Foundation (NWO) Planetary and Exoplanetary Science (PEPSci) grant. B.-O. D. acknowledges support from the Swiss State Secretariat for Education, Research and Innovation (SERI) under contract number MB22.00046.

**Author Contributions**

R.H. designed the observations, led the interpretation, and simulated the magma ocean - atmosphere models. A.B.-A. led the data analysis using Eureka!. M.Z. led the data analysis using SPARTA. K.P. and H.A.K. provided spectral retrievals. M.Z., C.B., and Y.M. provided self-consistent models for vaporized-rock and volatile atmospheres. M.B., Y.P., D.D., A.B., B.-O.D. provided independent data analyses. M.D. contributed to the design of the observations and the data analysis. Y.I. provided independent models of vaporized-rock atmospheres. M.S. developed the climate routine used for the magma ocean - atmosphere models. All authors contributed to the writing of the manuscript.

**Additional Information**

Reprints and permissions information is available at [www.nature.com/reprints](www.nature.com/reprints). The authors declare no competing interests. Correspondence and requests for materials should be addressed to Renyu Hu ([renyu.hu@jpl.nasa.gov](renyu.hu@jpl.nasa.gov)).

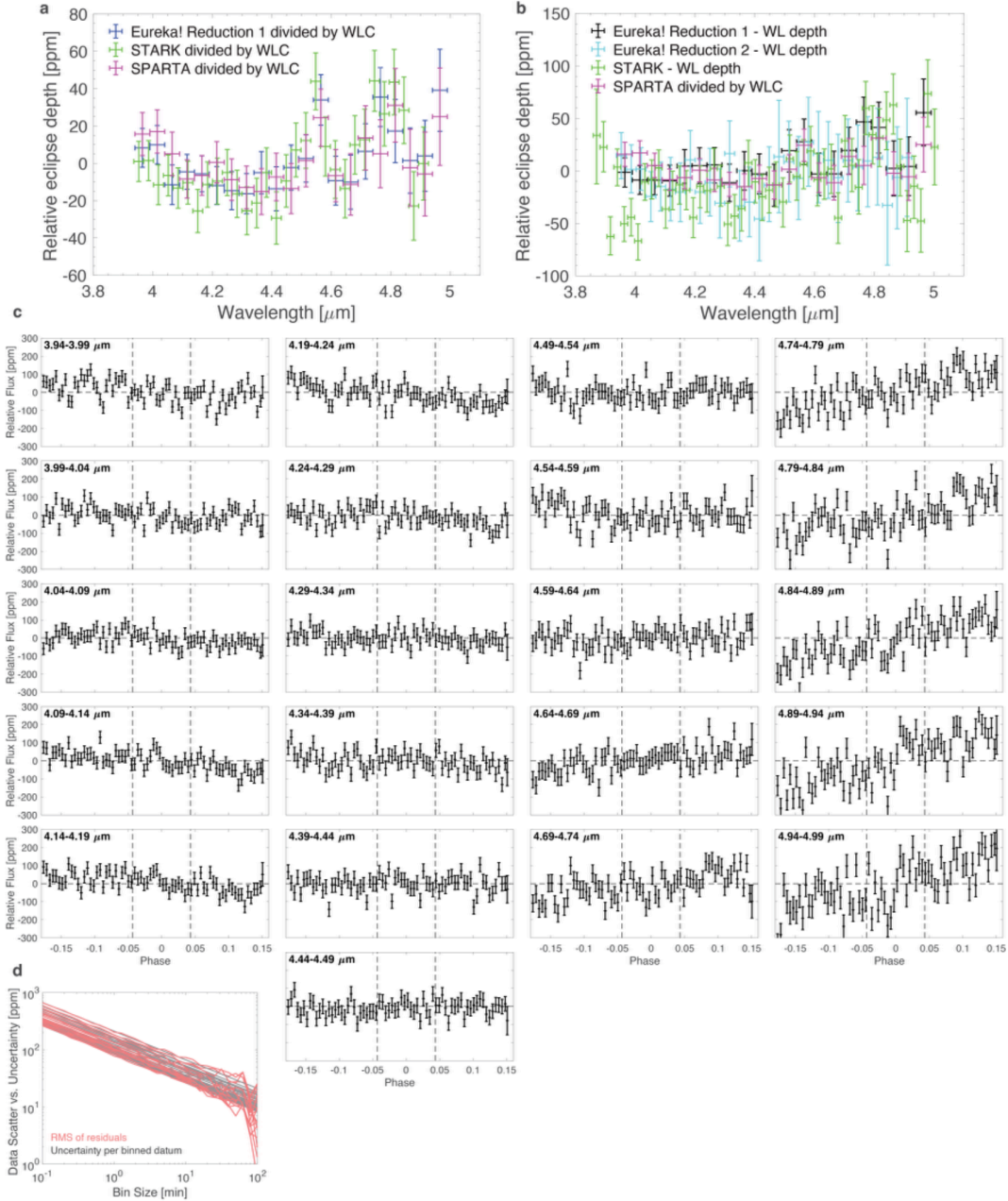

**Extended Data Fig. 1: Reduction of the NIRCam eclipse observation of 55 Cnc e. a,** The relative eclipse depths derived by fitting to the relative light curves obtained by dividing the spectroscopic light curves by the white light curve (WLC). The spectrum is consistent between the analyses, and we adopt the SPARTA analysis for model interpretation. **b,** The SPARTA analysis in comparison with the eclipse depths derived by direct fitting to the spectroscopic light curves (minus respective while-light (WL) depth). Direct fitting results in considerably noisier spectra, while the overall shape is approximately consistent with the SPARTA analysis. **c,** Relative spectroscopic light curves in the SPARTA analysis. The vertical lines denote the phases where the planet is eclipsed by the star. **d,** Comparison between the scatters of the light curves and the uncertainties of binned data, for the relative spectroscopic light curves from the SPARTA analysis. All error bars correspond to 1σ.

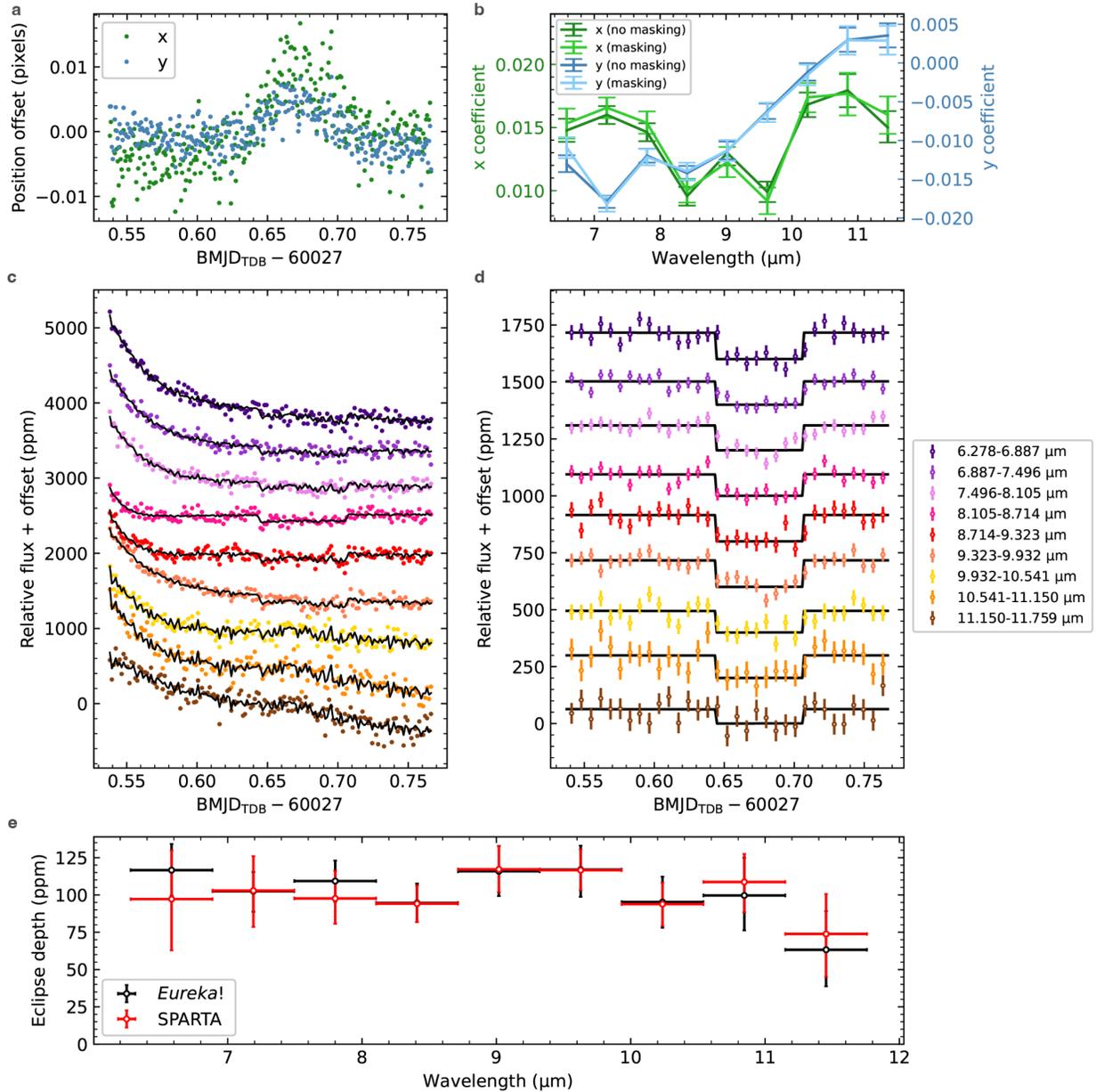

**Extended Data Fig. 2: Reduction of the MIRI eclipse observation of 55 Cnc e. a,** Drift of the trace along the spectral (x) and spatial (y) directions during the observation, binned to 1-minute intervals. **b,** Wavelength-dependent coefficients for linear decorrelation against the drift in the x and y directions, calculated with and without masking the in-eclipse data. **c,** Raw spectroscopic light curves and best-fit models, binned to 2-minute intervals for easier visualization, after trimming the first 30 minutes of data. **d,** Systematics-corrected spectroscopic light curves and eclipse models, binned to 10-min intervals. **e,** Comparison between the *Eureka!* and SPARTA eclipse depth spectra. The spectrum is consistent between the analyses, and the *Eureka!* analysis has smaller final uncertainties (due to less scatter in binned light curves) and is thus adopted for model interpretation. All error bars correspond to 1σ.

| Param | Value | Note |
|---|---|---|
| $R_p$ | 1.95±0.04 $R_{Earth}$ | Ref[12] |
| $M_p$ | 8.6±0.4 $M_{Earth}$ | Ref[12] |
| d | 12.59±0.01 pc | Gaia DR2 |
| $R_p/R_*$ | 0.0182±0.0002 | Ref[11] |
| $a/R_*$ | 3.52±0.01 | Ref[11] |
| $R_*$ | 0.98±0.02 $R_{sun}$ | Ref[12] |
| $M_*$ | 1.02±0.05 $M_{sun}$ | Ref[12] |
| $L_*$ | 0.6396±0.0009 $L_{sun}$ | Gaia DR2 |
| $T_{eff}$ | 5214±53 K | Derived from $L_*$ and $R_*$ |
| a | 0.0160±0.0003 AU | Derived from $a/R_*$ and $R_*$ |
| $T_{eq}$ | 1965±20 K | Derived for zero albedo and full thermal redistribution (f=1/4) |
| $T_{eq}$ | 2511±26 K | Derived for zero albedo and no thermal redistribution (f=2/3) |

**Extended Data Fig. 3: Parameters adopted in retrievals and self-consistent models.**

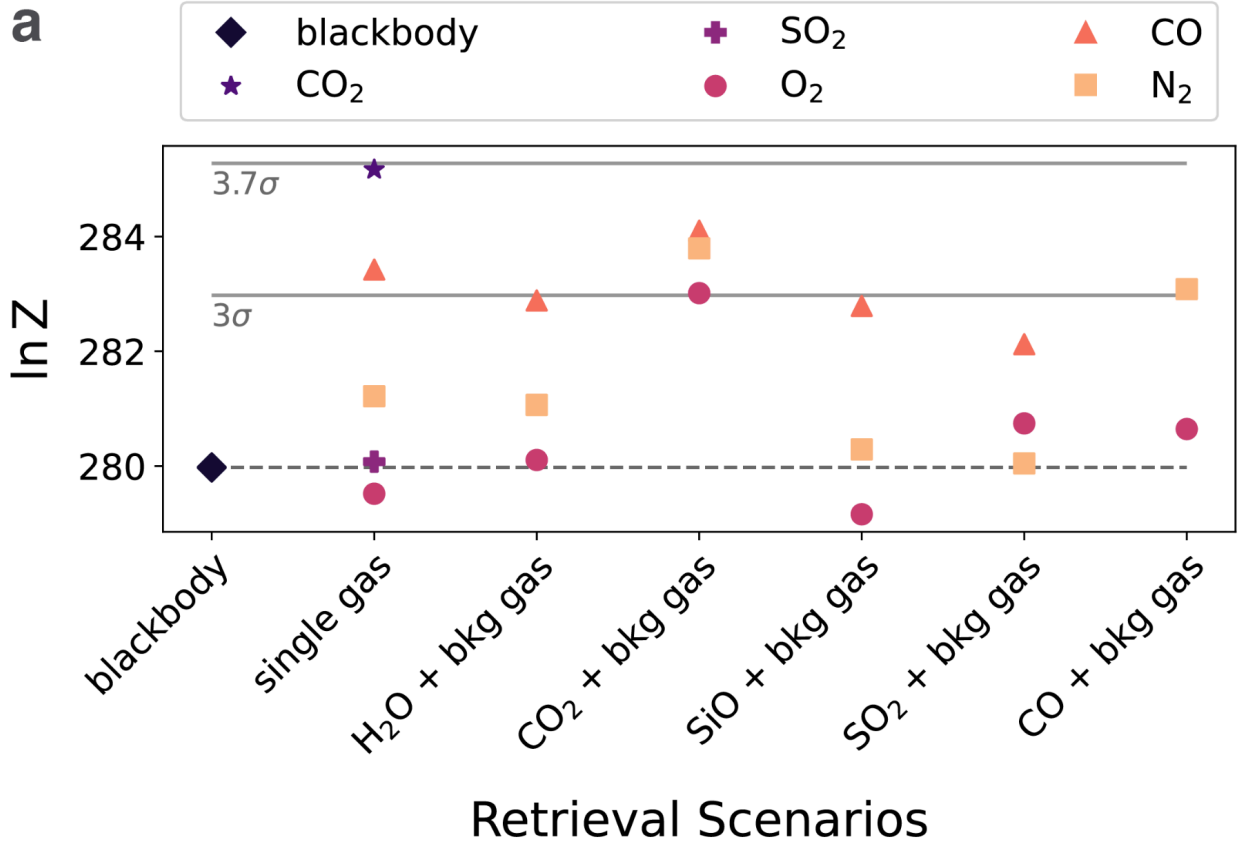

| Parameter | CO only | CO$_2$ only | CO$_2$ + CO | CO$_2$ + N$_2$ |
|---|---|---|---|---|
| lnZ | 283.4 | 285.2 | 284.1 | 283.8 |
| $\chi^2$/dof | 21.6/25 | 19.6/25 | 19.8/24 | 19.8/24 |
| NIRCam mean depth [ppm] | 64±8 | 63±7 | $63^{+8}_{-7}$ | $63^{+7}_{-8}$ |
| log10 ($P_{surf}$ [Pa]) | 6.2±1.0 | $3.7^{+2.4}_{-1.0}$ | $5.5^{+1.4}_{-1.7}$ | $5.3^{+1.6}_{-1.7}$ |
| $T_{surf}$ [K] | 1780±60 | $1880^{+80}_{-70}$ | 1835±80 | $1860^{+80}_{-70}$ |
| log10 (CO$_2$ VMR) | – | – | $-3.0^{+1.9}_{-3.4}$ | $-2.0^{+1.3}_{-1.9}$ |

**Extended Data Fig. 4: Summary of key retrieval results. a, Bayes evidence (Z) when fitting both the NIRCam and MIRI datasets.** The Bayes evidence was obtained by fitting 55 Cnc e's thermal emission spectrum with a blackbody and varied atmospheric scenarios in a spectral retrieval framework. **b, Qualities and parameter constraints from the preferred scenarios.** The uncertainties are 1σ.

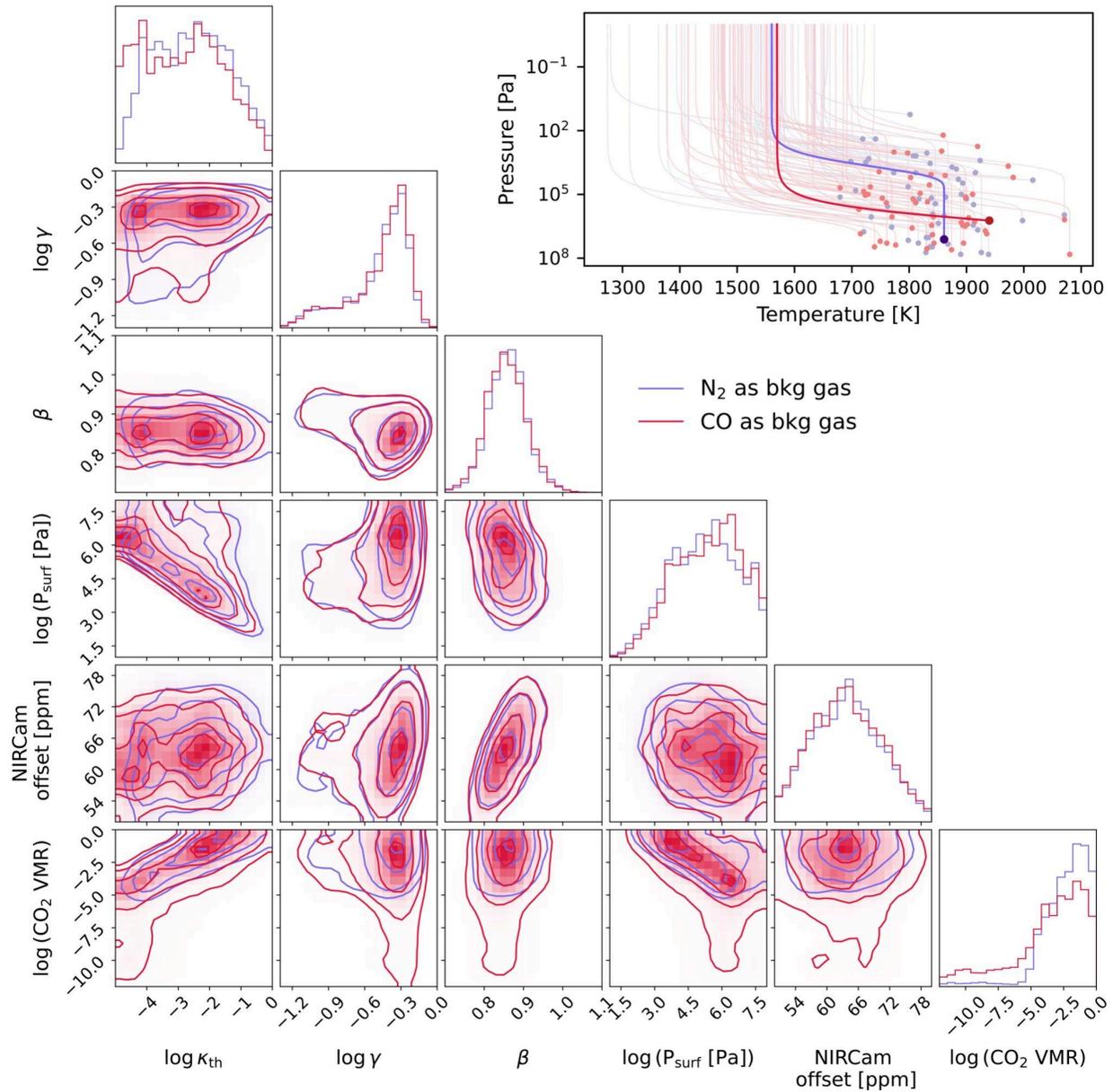

**Extended Data Fig. 5: Posterior distribution of retrieval parameters and sample pressure-temperature profiles.** The full posterior distribution of parameters and randomly sampled pressure-temperature profiles were obtained by fitting the emission spectrum to an $N_2$- (blue) or CO-dominated (red) atmosphere with varied volume mixing ratios of $CO_2$ and pressure-temperature profiles.

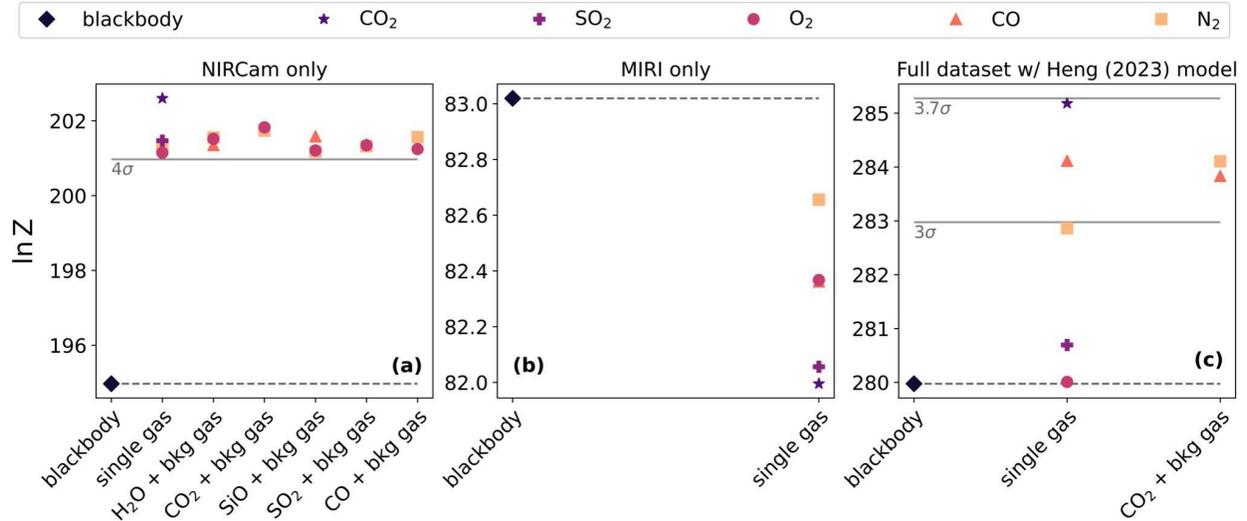
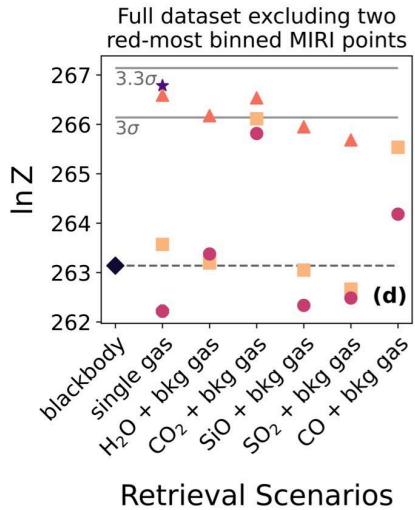
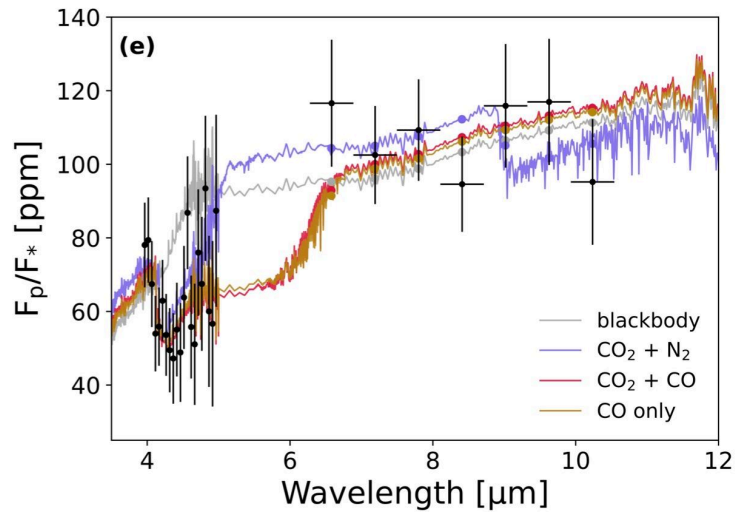

**Extended Data Fig. 6: Sensitivity to retrieval assumptions. a.** The Bayes evidence when fitting to the NIRCam data only. We allowed the mean eclipse depth to vary freely, and applied a physically motivated lower bound on the temperature (1700 K, corresponding to the equilibrium temperature with full heat redistribution and a high Bond albedo of 0.5). **b.** The Bayes evidence when fitting to the MIRI data only. **c.** The Bayes evidence when fitting to the full dataset (NIRCam + MIRI) with an independent parameterization of the temperature profile, where the surface temperature and the atmospheric temperature (assumed to be isothermal) determine the emission spectra[46]. **d.** The Bayes evidence when fitting to the full dataset excluding the two red-most binned MIRI data points (corresponding to the shadow region). **e.** The best-fit models when fitting to the full dataset excluding the two red-most binned MIRI data points. Colored points show the model results binned to the same wavelength channels as the MIRI data.

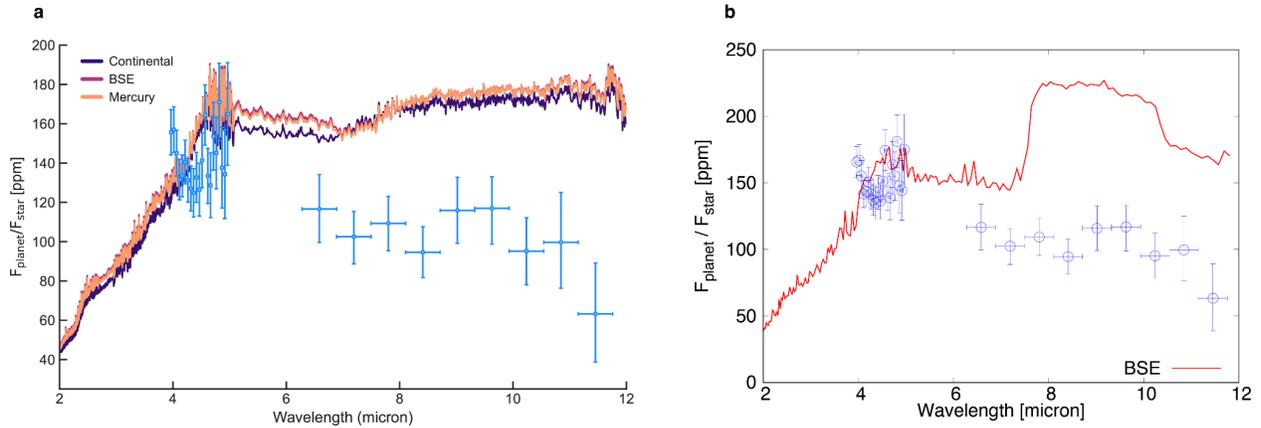

**Extended Data Fig. 7: Thermal emission spectra of 55 cnc e if it has a thin, vaporized-rock atmosphere. a.** The spectrum is calculated with the model of Ref[32] for varied silicate-based melt compositions. **b.** The spectrum is calculated with the model of Ref[31] assuming the BSE composition as magma composition. The NIRCam spectrum is shown with a mean eclipse depth of 150 ppm in both panels. The difference between the two models is mainly due to the different opacities used for SiO, but regardless a vaporized-rock atmosphere is inconsistent with the MIRI measured spectrum.

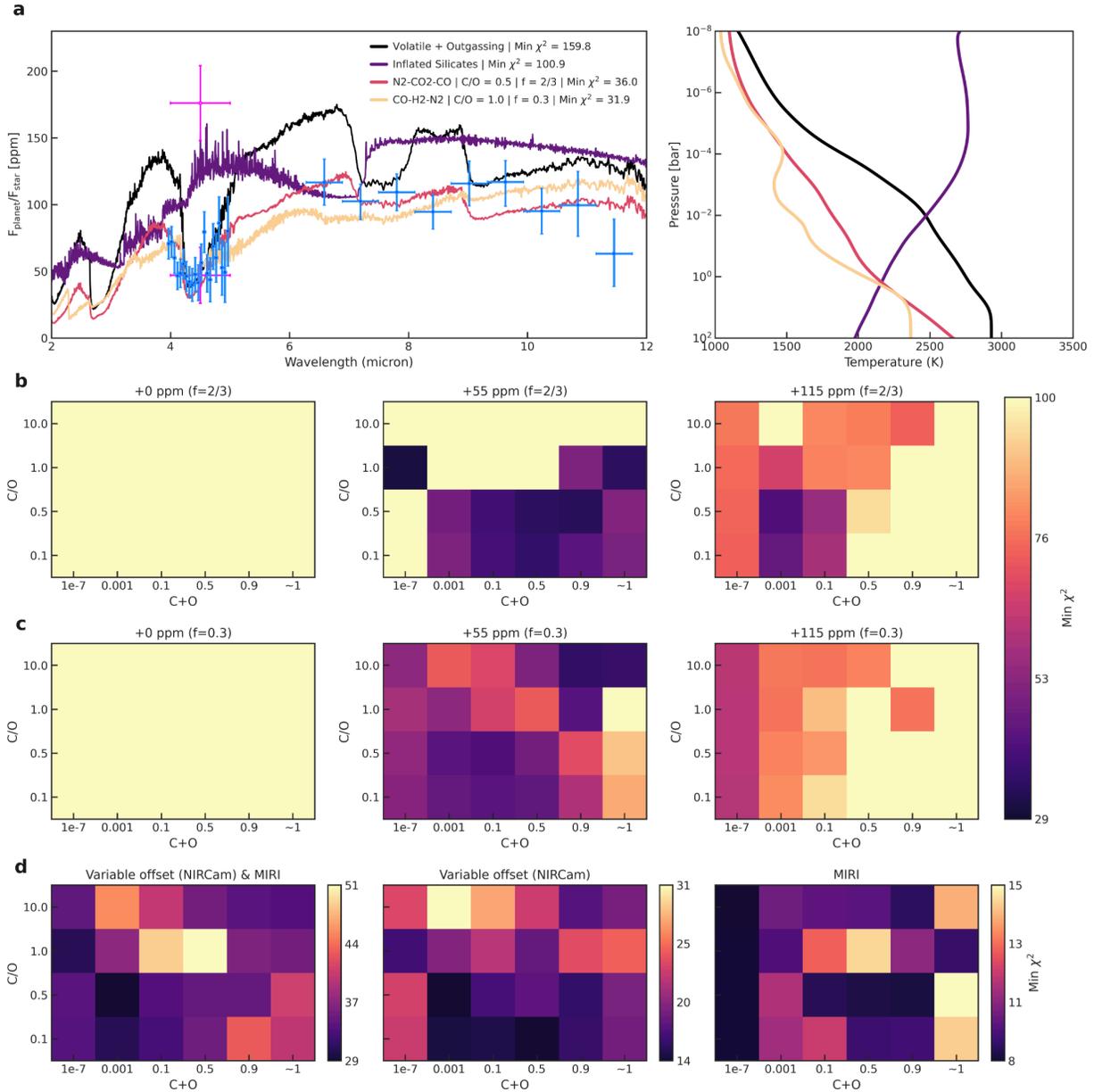

**Extended Data Fig. 8: Grid of self-consistent models for various volatile (CHONSP) compositions. a.** Emission spectra and temperature profiles of best-fit volatile-only models, compared with variant models enriched by rock-forming elements. The labels denote the dominant atmospheric constituents, as well as the C/O ratio and heat redistribution parameter. $\chi^2$ values represent the minimum achieved when allowing the NIRCam mean eclipse depth to vary freely, and the NIRCam data are shown with a mean eclipse depth of 55 ppm. The two magenta data points represent previous Spitzer observations[24]. **b.** $\chi^2$ values against the C+O mole fraction and the initial C/O ratio for three NIRCam mean eclipse depths for f=2/3 grid models. **c.** Same as panel b but for models with f=0.3. **d.** Minimum achieved $\chi^2$ values when the NIRCam mean depth is allowed to vary freely. The leftmost panel includes both NIRCam and MIRI data; the middle panel shows the minimum $\chi^2$ values when only fitting the NIRCam data; the rightmost panel shows the minimum $\chi^2$ values when only fitting the MIRI data.

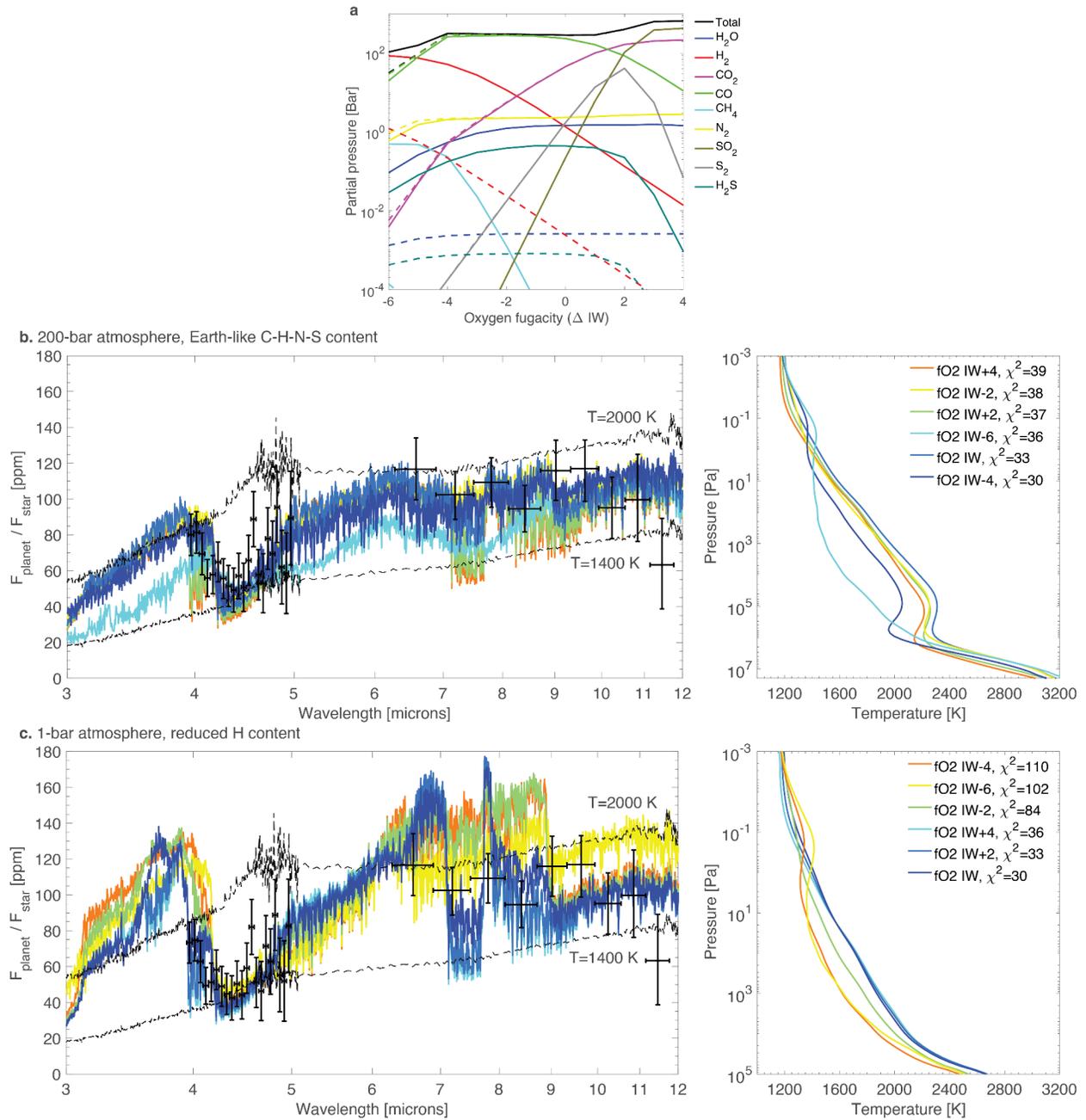

**Extended Data Fig. 9: Modeled atmosphere in equilibrium with a magma ocean on 55 Cnc e. a.** Partial pressure of gases for the volatile content of H, C, N, and S as Earth's early magma ocean (solid lines) and 3% H content (dashed lines), simulated by a magma ocean-atmosphere partitioning and chemical equilibrium/speciation model[44]. **b.** Emission spectra and associated pressure-temperature profiles for a 200-bar atmosphere with Earth-like C-H-N-S volatile abundance ratios in the atmosphere-magma ocean system, simulated by our atmospheric radiative-transfer and chemistry model. **c.** Same as panel b but for a 1-bar atmosphere and an H abundance ratio in the atmosphere-magma ocean system reduced to 3% of the standard cases. The NIRCam data are offset to match with the best-fit model (64 ppm for panel b and 58 ppm for panel c). The total number of data points is 30 for the interpretation of the $\chi^2$ values.

45